\documentclass[twocolumn]{aastex61}
\pdfoutput=3 
\usepackage{amsmath,amstext}
\usepackage[T1]{fontenc}
\usepackage{apjfonts} 
\usepackage[figure,figure*]{hypcap}
\usepackage{graphicx}
\usepackage{verbatim}

\submitjournal{ApJ}

\shorttitle{Discovery of Six Recycled Pulsars from AO327}
\shortauthors{Martinez et al.}

\begin{document}

\title{The Discovery of Six Recycled Pulsars from the Arecibo 327-MHz Drift-Scan Pulsar Survey}


\correspondingauthor{Jose Guadalupe Martinez}
\email{jmartinez@mpifr-bonn.mpg.de}

\author[0000-0003-0669-865X]{J. G. Martinez}
\affiliation{Max-Planck-Institut f\"{u}r Radioastronomie, Auf dem H\"{u}gel 69, D-53121 Bonn, Germany}
\author{P. Gentile}
\affiliation{Department of Physics and Astronomy, West Virginia University, 111 White Hall, Morgantown, WV 26506, USA}
\affiliation{Center for Gravitational Waves and Cosmology, West Virginia University, Chestnut Ridge Research Building, Morgantown, WV 26505, USA}
\author{P. C. C. Freire}
\affiliation{Max-Planck-Institut f\"{u}r Radioastronomie, Auf dem H\"{u}gel 69, D-53121 Bonn, Germany}
\author{K. Stovall}
\affiliation{National Radio Astronomy Observatory, P.O. Box 0, Socorro, NM 87801, USA}
\author{J. S. Deneva}
\affiliation{George Mason University, Resident at the Naval Research Laboratory, Washington, DC 20375, USA}
\author{G. Desvignes}
\affiliation{Max-Planck-Institut f\"{u}r Radioastronomie, Auf dem H\"{u}gel 69, D-53121 Bonn, Germany}
\author{F. A. Jenet}
\affiliation{Center for Advance Radio Astronomy, University of Texas at Rio Grande Valley, One West University Boulevard, Brownsville, TX 78520, USA}
\author{M. A. McLaughlin}
\affiliation{Department of Physics and Astronomy, West Virginia University, 111 White Hall, Morgantown, WV 26506, USA}
\affiliation{Center for Gravitational Waves and Cosmology, West Virginia University, Chestnut Ridge Research Building, Morgantown, WV 26505, USA}
\author{M. Bagchi}
\affiliation{The Institute of Mathematics Science (IMSc-HBNI), 4th Cross Road, CIT Campus Taramani, Chennai 600 113, India}
\author{Tom Devine}
\affiliation{Department of Physics and Astronomy, West Virginia University, 111 White Hall, Morgantown, WV 26506, USA}

\begin{abstract}
Recycled pulsars are old ($\gtrsim10^{8}$ yr) neutron stars that are descendants from close, interacting stellar systems. In order to understand their evolution and population, we must find and study the largest number possible of recycled pulsars in a way that is as unbiased as possible. In this work, we present the discovery and timing solutions of five recycled pulsars in binary systems (PSRs J0509$+$0856, J0709$+$0458, J0732$+$2314, J0824$+$0028, J2204$+$2700) and one isolated millisecond pulsar (PSR J0154$+$1833). These were found in data from the Arecibo 327-MHz Drift-Scan Pulsar Survey (AO327). 
All these pulsars have a low dispersion measure (DM) ($\lesssim 45 \, \rm{pc}\, cm^{-3}$), and have a DM-determined distance of $\lesssim$ 3 kpc.
Their timing solutions, have data spans ranging from 1 to $\sim$ 7 years,
include precise estimates of their spin and astrometric parameters, and for the binaries, precise estimates of their Keplerian binary parameters. Their orbital periods range from about 4 to 815 days and the minimum companion masses (assuming a pulsar mass of 1.4 $\rm{M_{\odot}}$) range from $\sim$ 0.06--1.11 $\rm{M_{\odot}}$. 
For two of the binaries we detect post-Keplerian parameters; in the case of PSR~J0709$+$0458 we measure the component masses but with a low precision, in the not too distant future the measurement of
the rate of advance of periastron and the Shapiro delay will
allow very precise mass measurements for this system. 
Like several other systems found in the AO327 data, PSRs J0509$+$0854, J0709$+$0458 and J0732$+$2314 
are now part of the NANOGrav timing array for gravitational wave detection.
\end{abstract}

\keywords{pulsars: general --- pulsars: individual:PSR J0709+0458 --- stars: neutron --- binaries: general}

\section{Introduction} \label{sec:intro}

\subsection{Recycled pulsars}

Pulsars that have undergone a binary interaction history are known as ``recycled pulsars''. These
neutron stars (NS) are the first formed compact object in a binary system. They were then spun
to high spin frequencies via accretion of mass and angular momentum from the secondary star
\citep{Alpar1982}; during this phase these systems are observable as X-ray binaries 
\citep{Tauris_van_den_Heuvel2006}.

Recycling seems to decrease the surface dipolar magnetic fields of these neutron stars to
relatively low values ($B \, \sim\,  10^{8} - 10^{10} \, \rm G$). These low values of $B$ result
in very small spin-down rates and, consequently, large ($\gtrsim10^{8} \, \rm yr$) characteristic ages.

Recycled pulsars can be further divided into two subclasses,  millisecond pulsars (MSPs), which
we define here 
as having spin periods smaller than that of the fastest-spinning pulsar in a double neutron star
system, 16.7 ms \citep{Stovall_J1946}, and mildly recycled pulsars (MRPs), which have larger spin periods.
The reason for the slower spin periods of MRPs is that (generally) the progenitors of their
companions were massive stars with faster evolution; for such systems the recycling phase is short-lived
\citep{Tauris_Ultra-stripSN}; these systems have smaller decreases in $B$.

In contrast, the companions to MSPs had lower mass progenitors, which evolve much more slowly.
In such cases, the recycling process is longer, which allows for greater amounts of mass and
angular momentum to be accreted onto the NS \citep{TaurisLangerKramer2011,TaurisLangerKramer2012} and
a more extensive reduction of the magnetic field. All recycled pulsars are relics of the evolution of
close, interacting binary systems, and their observed properties are ancient records of their
evolutionary history.

\subsection{Applications}

Recycled pulsars are stable rotators and can be timed very precisely. This property allows
many applications: using them as key probes of stellar astrophysics \citep{Bhattacharya&van_den_Heuvel1991,TaurisLangerKramer2011,TaurisLangerKramer2012},
tests of General Relativity (GR) using double neutron star (DNS) systems, such as
PSR~J0737$-$3039 \citep{GRTestsdoublePSR}, PSR~B1913$+$16 \citep{2016ApJ...829...55W},
PSR~J1946$+$2052 \citep{Stovall_J1946} and PSR~J1757$-$1854 \citep{Cameron_J1757}; tests of
the nature of gravitational waves and of
alternate theories of gravity using the orbital decay of MSP-white dwarf (WD) systems such as
PSRs~J1738$+$0333 and J0348$+$0432 \citep{Freire+12,Antoniadis+13_J0348}
and the universality of free fall in the triple system
PSR J0337$+$1715 \citep{Ransom+14_TripleSystem,Archibald+18_TripleSystem}.
They have set strong constraints on the equation of state of dense matter by extending the
known mass limits of a neutron star from systems like PSR~J0348$+$0432 \citep{Antoniadis+13_J0348}.
Last but foremost, recycled pulsars, more specifically MSPs, have helped set limits on the low-frequency
gravitational wave background in the Universe by using pulsar timing arrays
(PTAs)\citep{GWB_limit_NANOGrav,GWB_limit_PPTA,GWB_limit_EPTA}, with sensitivities expected to grow in time.

\subsection{The Arecibo 327-MHz Drift-Scan Pulsar Survey}

Given this vast range of applications, many ongoing large scale pulsar surveys are currently being undertaken with the objective of finding new millisecond pulsars. One of these is
the Arecibo 327-MHz Drift-Scan Pulsar Survey (AO327).
This survey and its strategy have been
described in detail by  \citep{Deneva13}. Briefly, the AO327 will search for pulsars
in the entire Arecibo sky (declination range from $-1^{\circ}$ to $38^{\circ}$) in drift-scan
mode at a center frequency of 327-MHz. In this mode, the telescope beam is pointed at the meridian
at some fixed declination, and the back-end records the signal as the Earth rotates the radio telescope
and its beam. Given the size of the telescope beam ($\sim \, 15 \arcmin$), a pulsar at the same
declination will be within the beam approximately 1 minute (slightly more at the higher declinations).
Despite the small observation time, the survey is still very sensitive to new pulsars because of the
large collecting area of the telescope and the large fractional bandwidth of the receiver.
The short observation time makes the survey sensitive to binary pulsars with short (10 to 93 minute)
orbital periods, which have not been discovered to date at radio wavelengths.

The survey began in 2010 and since then it has a running total of 85 discoveries,
including 16 new recycled pulsars, 10 of which are MSPs, and 16 rotating radio transients (RRATs).
The recycled pulsars include two noteworthy
DNS systems: PSR~J0453$+$1559, the first asymmetric DNS with the smallest precisely
measured mass of any neutron star ~\citep{Martinez_J0453+1559} and PSR~J1411$+$2551, one of the
lowest total mass DNSs known \citep{Martinez_J1411+2551}. This survey also discovered PSR~J2234$+$0611,
a MSP with an eccentric orbit with high timing precision and optical spectroscopic
measurements of the Helium WD companion, which make this binary system a great laboratory for
studying stellar evolution \citep{Deneva13,J2234_optical,2019ApJ...870...74S} and has, together with
several of the systems presented in this work, been included in the NANOGrav PTA. 

\subsection{Motivation and structure of this paper}

In this letter, we present the discovery and follow-up timing of six recycled pulsars found
in AO327 data:
four MSPs (PSRs J0154$+$1833, J0509$+$0856, J0732$+$2314, and J0824$+$0028) and two MRPs (PSRs J0709$+$0458 and J2204+2700). In Section~\ref{sec:timingObs}, we describe the observations used to discover and time these systems. In Section~\ref{sec:results}, we describe the details of each pulsar and the nature of binary companions. In Section~\ref{sec:postPK}, we describe the measurements of Post-Keplerian parameters
and their implications. We summarize our findings in Section~\ref{sec:conclusion}. 

\begin{figure*}\label{fig:prof1}
\centering
\includegraphics[scale=0.33]{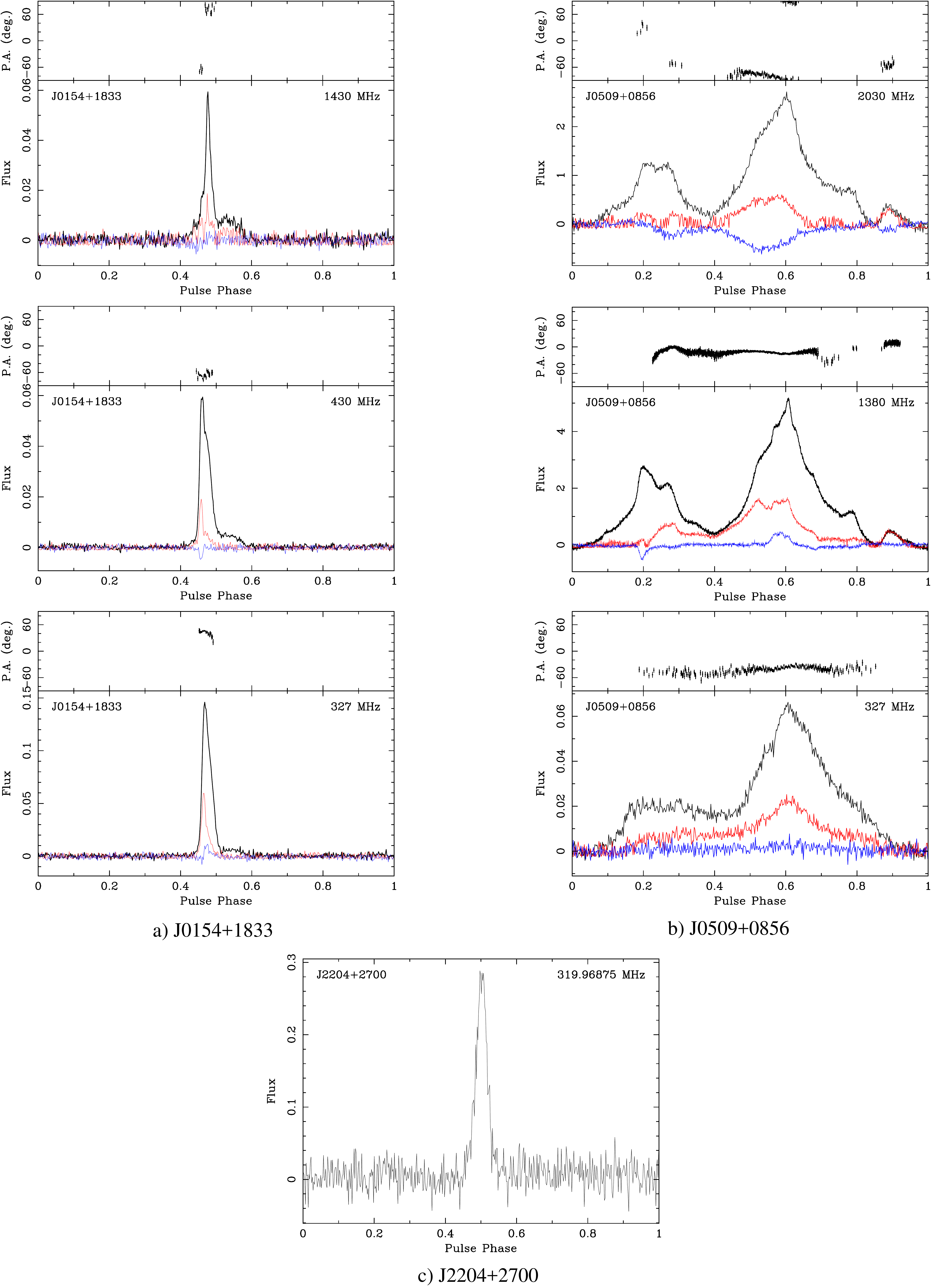}
\caption{Polarization calibrated pulse profiles for PSRs J0154$+$1833 and J0509$+$0856, based on full Stokes data taken at the radio frequencies shown on the right of each profile. They are obtained by averaging the best detections; the black line indicates the total intensity, the red line is the amplitude of linear polarization, and the blue line is the amplitude of the circular polarization. At the top of each polarization calibrated pulse profile, we show the position angle of the linear polarization.  At the bottom of this figure we show a pulse profile of PSR J2204$+$2700 showing only the total intensity in black. This pulsar is only strongly detected at 327-MHz but was observed for short observations due to its low timing precision, therefore we did not have enough signal to noise for polarization calibration. The pulse profile flux is in arbitrary units. }
\end{figure*}

\begin{figure}\label{fig:prof2}
\centering
\includegraphics[width=\columnwidth]{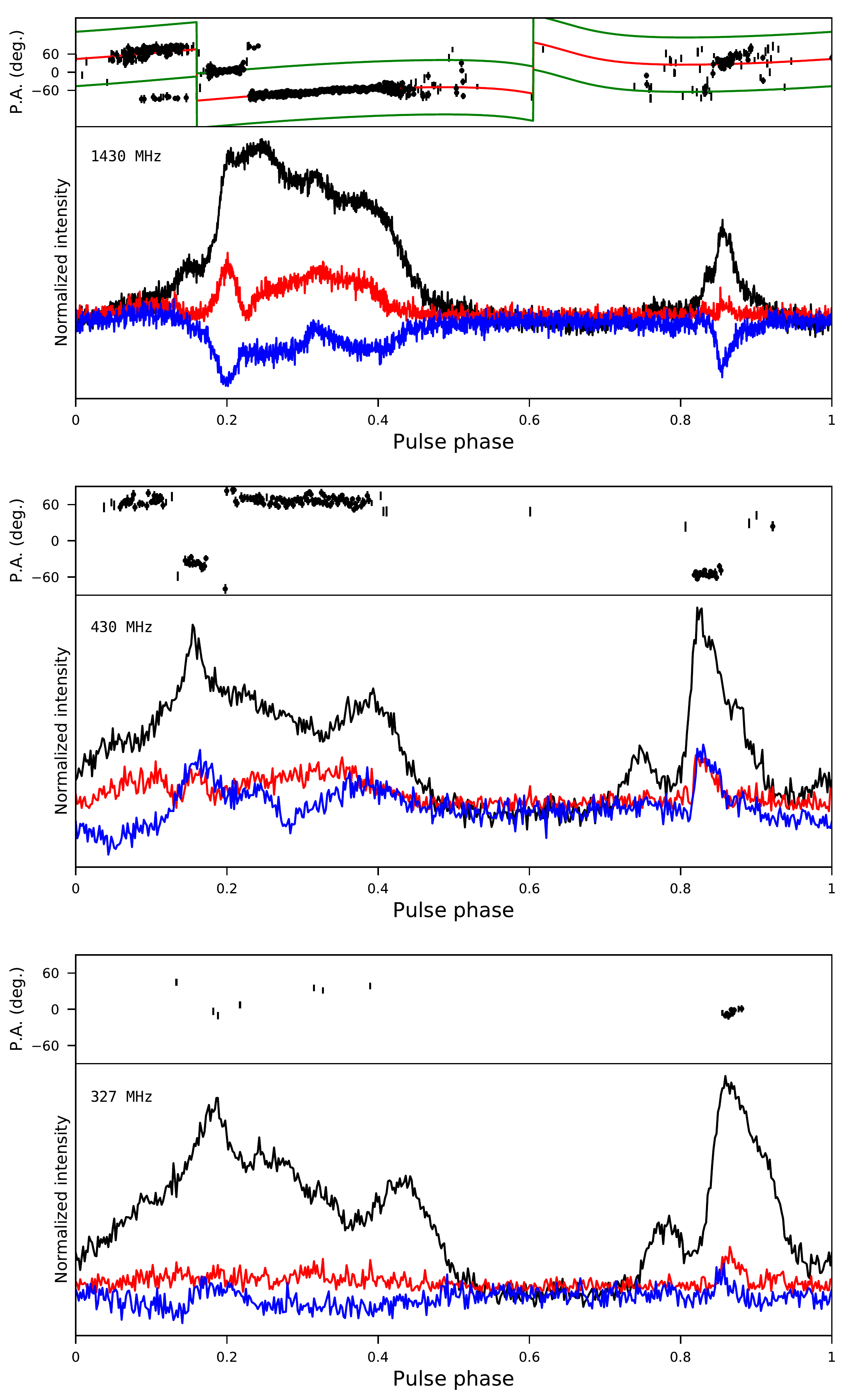}
\caption{Same as Figure~\ref{fig:prof1}, this time for PSR J0732$+$2314. For this pulsar, we also
display a rotating vector model at the top panel,  which is based on the L-band data (more details are in Section 3.4). The red line is the rotating vector model fit to the position angle measurements of this pulsar, while the green lines are a 90$^{\circ}$ orthogonal mode transitions.  }
\end{figure}

\begin{figure*}\label{fig:prof3}
\centering
\includegraphics[scale=0.35]{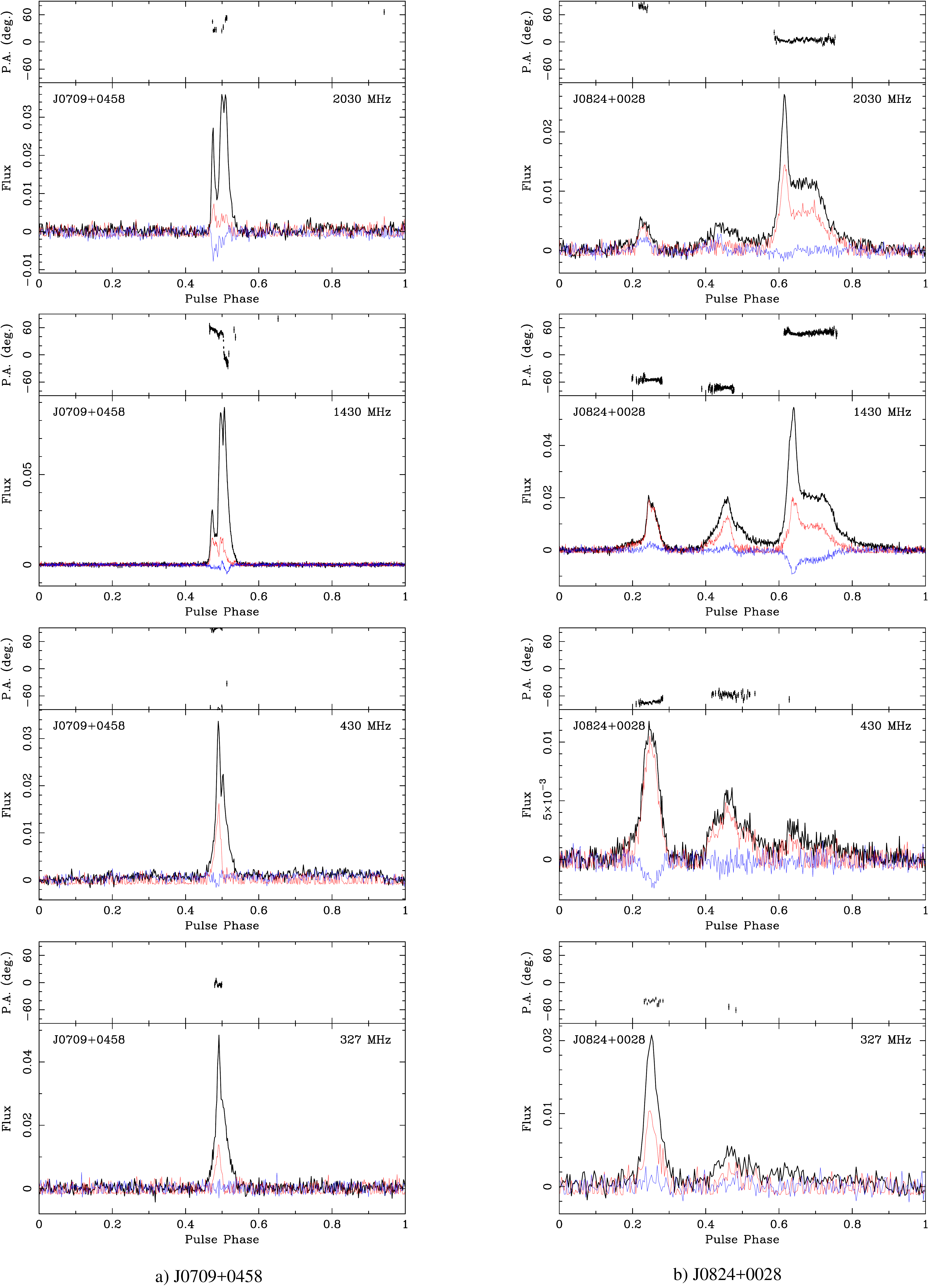}
\caption{Polarization calibrated pulse profiles for PSRs J0709$+$0458 and J0824$+$0028, based on full Stokes data taken at frequencies shown on the right of each profile. They are obtained by averaging the best detections; the black line indicates the total intensity, the red line is the amplitude of linear polarization, and the blue line is the amplitude of the circular polarization. In the top of each polarization calibrated pulse profile, we show the position angle of the linear polarization. The pulse profile flux is in arbitrary units.}
\end{figure*}

\section{Timing Observations and analysis} \label{sec:timingObs}

\subsection{Observations and data reduction}

The observations and the initial follow up of each pulsar discovered by AO327 were made
with the Arecibo 305-m radio telescope and the 327-MHz receiver, with the
 Puerto Rico Ultimate Pulsar Processing Instrument (PUPPI) as a back-end.
After the discovery of each pulsar (which is listed almost immediately on the AO327 website\footnote{http://naic.edu/~deneva/drift-search/}) we begin regular timing observations.
For the initial timing, PUPPI is used in incoherent search mode with a bandwidth of 68.7 MHz that is split in 2816 channels with a sample time of 82 $\mu$s. 
These are processed with the \texttt{PRESTO}\footnote{https://www.cv.nrao.edu/~sransom/presto/} pulsar software, first to remove radio frequency interference in the data, and then to dedisperse and fold the data at optimal spin periods and DMs.

When a pulsar is in a binary system, there is a Doppler modulation of the observed spin period that is 
caused by the orbital motion. By having sufficient data, one can fit a Keplerian model of the
orbit to the observed spin periods. 
This consists of the pulsar's spin period plus 5 orbital parameters: the orbital period ($P_b$), the
semi-major axis of the pulsar's orbit projected along the line of sight, normally denominated in light
seconds ($x$), the orbital eccentricity ($e$), the longitude of periastron ($\omega$) and the time of
passage through periastron ($T_0$). To do this we use  \texttt{FITORBIT}~\footnote{http://www.jb.man.ac.uk/pulsar/Resources/tools.html}.

Using this Keplerian model, we then dedisperse and fold all existing data for the pulsar, this results
in clear pulse profiles.
We then calculate the pulsar's pulse times of arrival (TOAs) via Fourier-domain cross-correlation of these
profiles with a noise-free template and the observed pulse profile \citep{TaylorRG}.
These templates were created by fitting one or more Gaussians to the pulse profiles but for the complicated shape profiles we averaged the best detection  with low-noise observations together for a template.

We then use the TOAs to determine the phase coherent timing solutions for the pulsars,
using the manual technique described in section 3 of \cite{connect}.
For this, we use \texttt{TEMPO}\footnote{\url{http://tempo.sourceforge.net}}, a pulsar timing program, to fit a timing model to the observed TOAs. This model consists of spin frequency and its derivative, position, and dispersion measure (DM) and the Keplerian orbital parameters mentioned above, all determined
to much higher precision than possible from the analysis of the spin periods.
If necessary, post-Keplerian orbital parameters are also included.

With these solutions, we can account for every single rotation of the pulsar
for the whole observational data span.
In the case of PSR~J0732$+$2314, the data from the densest timing campaign
was not dense and precise enough to obtain an unambiguous phase connection using 
the basic phase-connection technique mentioned above.
This problem happens frequently for pulsars with relatively wide orbits, where
the astrometric, spin and orbital timescales are similar. In order to solve these systems,
generally several sets of relatively dense observations are necessary.
Instead of requesting more observations, we used the algorithm described
in section 4 of \cite{connect} to finally obtain unambiguous phase connection.

After obtaining these solutions, we were able to use PUPPI's coherent folding mode for these
pulsars. This mode coherently dedisperses and folds the data online, completely removing the dispersive
(but not the scattering) effects of the interstellar medium. Observations taken in this mode result
in much improved timing precision and much smaller data rates.
The precise position contained in the timing solution also means that we can use receivers other than the 327-MHz that have smaller beams, in particular the 430-MHz, L-wide and S-low. The 327-MHz and 430-MHz receivers have a total bandwidth of 87.5MHz (each have a visible bandwidth of 50 MHz and 25 MHz respectively) that is split in 56 channels. The L-wide and S-low receivers are sensitive to radio waves between 1150 and 1730 MHz and 1800 to 3100 MHz
respectively, we cover these bands (or the lower frequency part of them) with the 800 MHz bandwidth of PUPPI,
which is split in 512 channels. 

Most of the pulsars were observed at these 4 frequencies to determine an optimal
frequency for timing, defined as the frequency where the TOA rms is smallest. This
becomes the main frequency for observations of that pulsar. These multi-frequency observations were also
made in order to measure the polarimetric properties of the emission of these pulsars
at a range of frequencies; these are shown in Figures~\ref{fig:prof1} to \ref{fig:prof3}. Figure~\ref{fig:prof1}, additionally contains the total intensity, single frequency profile of PSR J2204$+$2700. Three of the pulsars presented in this work have broad, complex profiles
with many different components and emission from most of the spin cycle;
the components vary in shape and intensity at different frequencies,
particularly for PSRs~J0509$+$0856 (Figure~\ref{fig:prof1}), J0732$+$2314
(Figure~\ref{fig:prof2}) and J0824$+$0028 (Figure~\ref{fig:prof3}). All of the coherent folding mode observations made it possible to determine at a high precision each pulsar's rotational, astrometric, and if in a binary, orbital parameters.

\begin{figure*}\label{fig:residuals_epoch}
\centering
\includegraphics[scale=0.50]{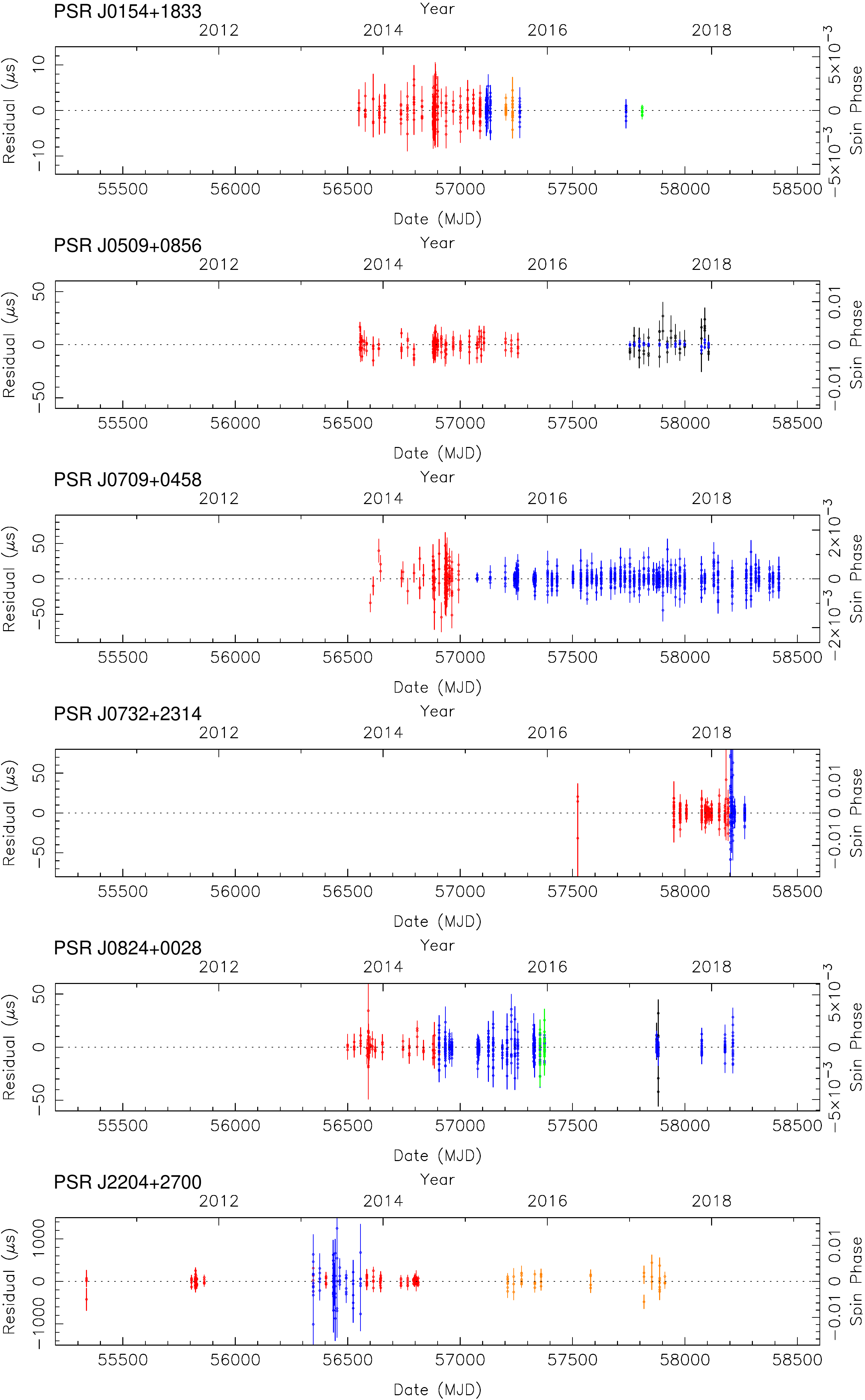}
\caption{TOA residuals as a function of epoch for the six recycled pulsars presented in this paper, with
1-$\sigma$ error bars.
In this and the following picture, the colors indicate different frequencies and observation modes.
{\em Red:} Incoherent 327-MHz
data, {\em Orange}: coherent 327-MHz data, {\em Green}: coherent 430-MHz data, {\em Blue}: coherent
L-band (1.15-1.73 GHz) data and {\em Black}: coherent S-band (1.7-2.2 GHz) data. No unmodeled trends
are apparent in this figure and the next, suggesting that the timing solutions in Tables~\ref{timingSol_1} and \ref{timingSol}
provide a good description of the TOAs.}
\end{figure*}

\begin{figure*}\label{fig:residuals_orbital_phase}
\centering
\includegraphics[scale=0.50]{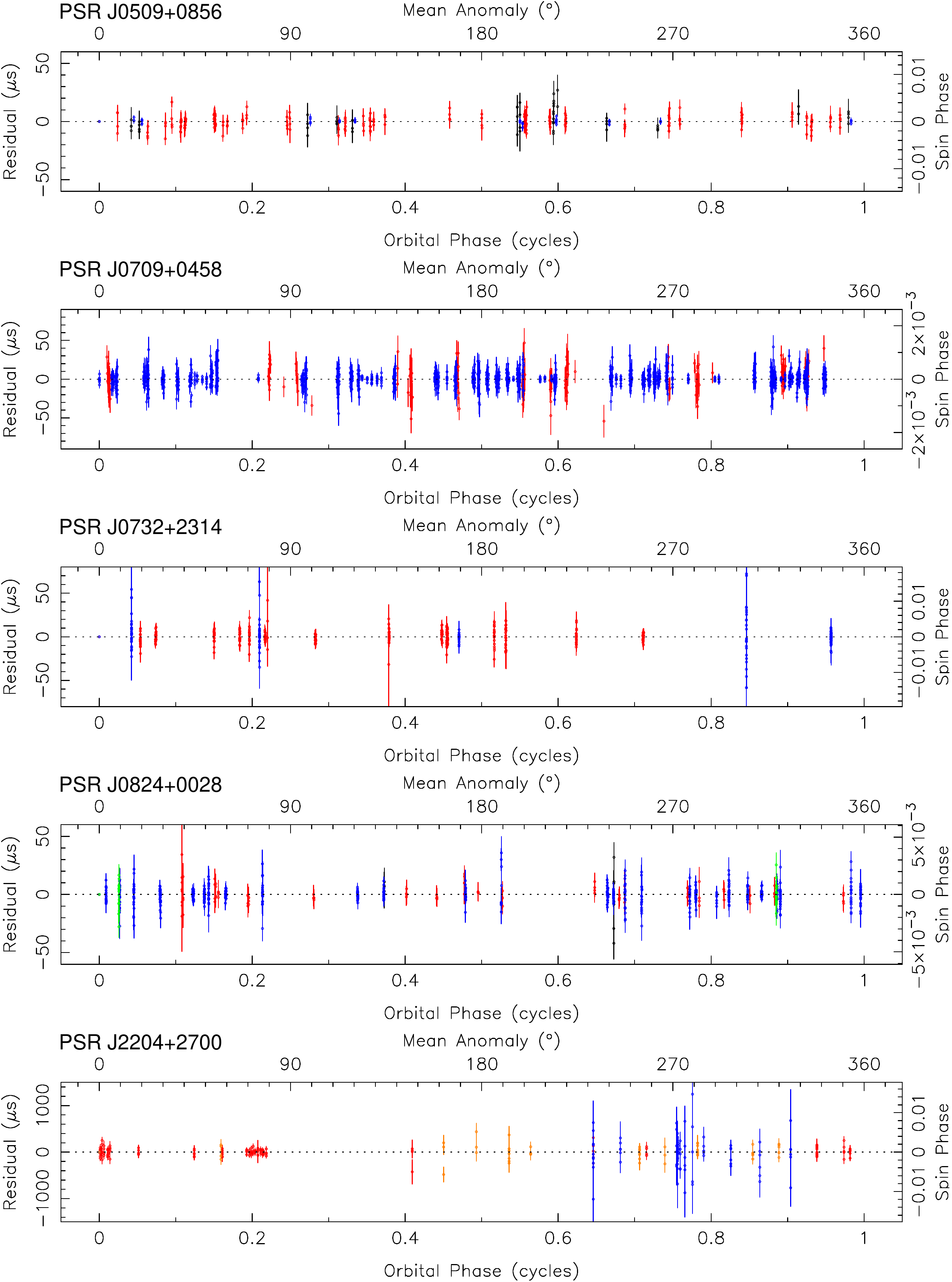}
\caption{Same as in Figure~\ref{fig:residuals_epoch}, but now with the residuals displayed as a function
of orbital phase for the five binary pulsars presented in this work. For PSRs~J0509$+$0856 and J0732+2314, where we used the ELL1 model, the mean anomaly is defined relative to the ascending node; for the other pulsars, the mean anomaly is defined relative to periastron.}
\end{figure*}

\subsection{Detailed timing analysis}

Finally, at the last stage of the analysis, all data is carefully re-folded and dedispersed
using the fully coherent timing solution, in order to eliminate any smearing of the pulse profiles.
The data was calibrated using the noise diode observations taken before each observation. Each observation is then corrected for the Faraday effect, which requires the measurement of the rotation measure (RM) by using \texttt{rmfit} from \texttt{PSRCHIVE} software \citep{PSRCHIVE,PSRCHIVE2012}.  At each epoch we generally split the data into four sub-integrations and eight frequency sub-bands, sometimes less due to short observations. Then we averaged the data in time and frequency, calculating an average of TOAs per epoch. All this results in much improved TOAs.

At the last step, we fit for the binary parameters we use two theory-independent timing models
based on the \cite{Damour86} model: these are the ELL1H and DDFWHE models.
The former was developed specifically for low-eccentricity binaries \citep{ELL1};
the Keplerian parameters $e$, $\omega$ and $T_0$ are replaced by
$\epsilon_1 \, \equiv \, e \sin \omega$, $\epsilon_2 \, \equiv \, e \cos \omega$ and the
time of ascending node, $T_{\rm asc}$. The advantage of these parameters is that is that they eliminate
the strong correlation between $\omega$ and $T_0$ that inevitably happens for low-eccentricity orbits;
unlike $T_0$, $T_{\rm asc}$ can be determined very precisely in such orbits.
The timing solutions derived from this model are presented in Table~\ref{timingSol_1}.

The cost of using the ELL1 model is that it assumes that the orbital eccentricity causes
a simple sinusoidal effect in the TOAs (with linear amplitude $x e$ and period $P_b / 2$), ignoring 
higher order contributions in $e$.
For this reason, we only use it if the ignored part is not relevant for the timing, i.e., if 
\begin{equation}
\label{eq:ecc}
x e^2 < \frac{\sigma}{\sqrt{N}},
\end{equation}
where $\sigma$ is the residual rms and $N$ is the number of TOAs.
If the condition given by eq.~\ref{eq:ecc} is not fulfilled, then those higher-order terms 
can in principle be detected by the timing, potentially having an effect on the measurement
of PK parameters, like the Shapiro delay. For such binaries we use the exact DDFWHE model;
which uses the Keplerian parameters $T_0$, $e$ and $\omega$.
Those timing solutions are presented in Table~\ref{timingSol}.
Regarding the Shapiro delay, both models use the orthometric parameterization \citep{FreireWex}.

We used the DE421 solar system ephemeris in all of the pulsar's timing models with the UTC(NIST) clock in dynamic barycenter time (TDB) units.
There are newer solar system ephemerides, however, detailed analyses by the NANOGrav consortium
reveal that no Solar System ephemeris is completely satisfactory
~\citep{NANOGRAV_11yr+18}. However, the differences between the Solar System ephemerides are
very small, resulting in differences
in the timing parameters of the pulsars that are much smaller than their estimated uncertainties. 

Four of the systems presented here (PSRs J0509+0856, J0709+0458, J0732+2314 and J0824+0028)
were timed by the NANOGrav, with the idea of verifying their timing stability and suitability for
pulsar timing arrays. For this reason, a significant amount of the timing data presented here was
taken by the NANOGrav collaboration, who share such data with the discoverers.

\section{Results} \label{sec:results}

All the final measured parameters of the pulsar's spin frequency and its derivative, position
and proper motion,
DM and, when appropriate, binary parameters are presented in Tables~\ref{timingSol_1} and \ref{timingSol}.
We also present in these tables some derived quantities. First, we derived the Galactic
coordinates ($l$ and $b$); then from these and the DMs we estimate the distances ($d$)
using NE2001 \citep{CordesLazio2002} and YMW \citep{YMW_model2017} models for the Galactic
distribution of free electrons. We also present the total proper motion, $\mu$ and
the transverse helocentric velocity.
Then, using the NE2001 distances we estimate the value of the
intrinsic spin period derivative ($\dot{P}_{\rm int}$) using
\begin{equation}
\dot{P}_{\rm int}\, = \, \dot{P}\,  - \, \frac{P}{c} \left( \mu^2 d \, + \, a_l  \right),
\end{equation}
where the first term in parentheses consist of the Shklovskii effect
\citep{Shklovskii70}, which is caused by the proper motion, and the second term
is the effect of the difference in the Galactic accelerations of the pulsar's system
and the Solar System projected along the direction from the pulsar to the Earth \citep{1991ApJ...366..501D}. In order to estimate $a_l$, we use the expressions presented by \cite{2009MNRAS.400..805L}, as the equation for the vertical acceleration should be valid to a Galactic height of $\sim \, \pm 1.5$ kpc, which is certainly the case for all the new systems
presented here. Additionally, for the acceleration parallel to the Galactic plane, we use the distance to the center of the Galaxy measured by the GRAVITY experiment \citep{2018A&A...615L..15G}, $r_0 \, = \, 8.122(31)$ kpc and a revised value for the rotational velocity of the Galaxy derived using the
latter $r_0$ \citep{2018arXiv180809435M}, $v_{\rm Gal} \, = \, 233.3$ km s$^{-1}$. 
The uncertainties on $\dot{P}_{\rm int}$ are derived by re-calculating it for distances
that are $\pm \, 25$\% of the NE2001 distance, i.e, they assume a small contribution
to the error from uncertainties in the proper motion. From $\dot{P}_{\rm int}$ we derive
the characteristic ages, $\tau_{c}$, inferred surface magnetic
field strengths, $B$, and spin-down energy loss rates, $\dot{E}$.
Finally, we present, for the binary systems, some derived binary parameters.
The minimum companion masses are derived using $M_{\rm p} \, = \, 1.4\, M_{\odot}$.

\begin{figure*}
\plotone{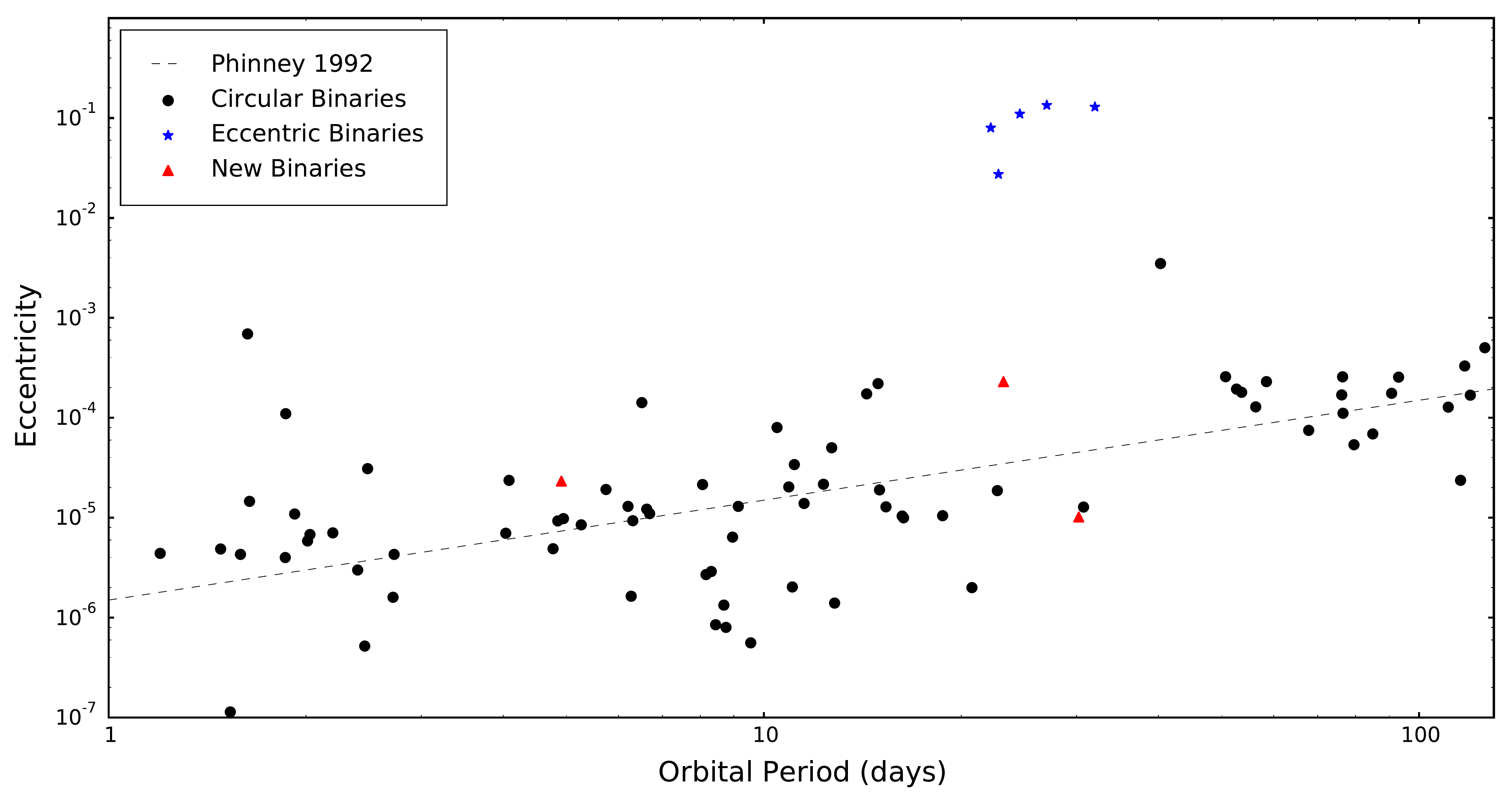}
\caption{Eccentricity ($e$) as a a function of orbital period ($P_{b}$) for recycled pulsars with low-mass companion ($<0.4 \rm{M_{\odot}}$). The dashed line represents the prediction expected from the evolutionary scenario by \citet{Phinney92}. Most recycled pulsars follow this line (shown as solid black circles), although eccentric ($e \sim 0.027 -0.14$, shown as in blue stars) binaries are found \citep{2015ApJ...806..140K,2015ApJ...810...85C,Barr2017,2018A&A...612A..78O,2019ApJ...870...74S} that likely result from a different evolutionary process \citep{Antoniadis14_formationEccenMSP}.  For some time, these systems were the only ones found in the $P_{b} \sim 22 - 32$ days region. The newly discovered PSRs J0732$+$2314 and J0824$+$0028 (shown as solid red triangles) are in that $P_{b}$ region, but are low eccentricity. Also PSR J0509$+$0856 follows \citet{Phinney92} evolutionary prediction very well. 
\label{fig:Eccen_vs_Pb}}
\end{figure*}

\subsection{PSR J0154$+$1833}
PSR J0154$+$1833 is isolated and the most rapidly rotating of the six recycled pulsars, with a spin period of 2.36 ms and a DM of 19.79 $\rm{pc}\, cm^{-3}$.  Our timing solution has a root mean square (RMS) residual of 1.23 $\mu$s and spans about 3 years. This pulsar's optimal timing frequency is at 327-MHz. Since most of the data for this pulsar was taken at 327-MHz, we retained great sensitivity to DM variations. This is important because this pulsar has a relatively
low ecliptic latitude ($\lambda_\odot\, =\, 33.168, \deg$, $\beta_\odot\, = \, 6.389 \deg$), which means that once a year it has a small angular separation with the Sun, and during this time the effects of the solar wind produce important deviations in the TOAs. We do not have enough data to model the effect of the solar wind in detail, which is modelled using two DM derivatives.

\subsection{PSR J0509$+$0856}
PSR J0509$+$0856 is an MSP with a spin period of 4.05 ms and a DM of 38.25 $\rm{pc}\, cm^{-3}$. It is in a 4.90 day almost circular ($e = 0.000023$) orbit. The minimum and median mass of the companion can be calculated from the measured mass function for this system. If we assume an orbital inclination of $i = 90^{\circ}$ and a pulsar mass of $1.4 \rm{M_{\odot}}$, we get a minimum companion mass of 0.11 $\rm{M_{\odot}}$ and a median companion mass (i.e. $i=60$) of 0.13 $\rm{M_{\odot}}$. 

The minimum companion mass is much lower than the prediction of the \cite{TS99} and
\cite{TaurisLangerKramer2012} models for the mass of a Helium WD companion for the
orbital period of this pulsar, which is of the order of $0.22\, \rm{M_{\odot}}$. This suggests
that, if the companion is a He WD (the spin period of the pulsar and the orbital eccentricity
of the system certainly are compatible with those of other MSP - He WD systems with similar orbital periods),
then either the system has a low orbital inclination (the 
inclination will be even lower if the companion is a CO WD), or, if orbital inclination is high,
then the pulsar mass is also high. The non-detection of the Shapiro delay in this system suggests the first
possibility. The polarimetric properties seem to confirm this: since the spin axis of the pulsar is nearly aligned
with the orbital plane, a face-on configuration should be one where the spin axis of the pulsar is nearly aligned 
with the Earth. In such a case, we should see a very broad profile, which we do observe in this system; indeed we
observe radio emission throughout nearly the full spin cycle. Also, the position angle of the linear polarization of
the system should vary slowly with spin phase, which is also observed in this system at the three radio frequencies
where we have measurements (Figure~\ref{fig:prof1}).

Our timing solution has an RMS residual of 3.64 $\mu $s and spans 4.8 years. This pulsar has very good timing precision
and has been regularly observed to determine its suitability for inclusion in the timing campaign being carried out by
NANOGrav, which took most of the data used in this work for this pulsar.

\subsection{PSR J0709$+$0458}

PSR J0709$+$0458 is a MRP that has a spin period of 34.42 ms and a DM of 44.26 $\rm{pc}\, cm^{-3}$.
It is in a 4.36-day binary orbit with an eccentricity of 0.000225. This system has a measurement, of the rate of advance of periastron, $\dot{\omega} = 0.032 \pm 0.012 ^{\circ}~{\rm yr}^{-1}$, and the orthometric Shapiro delay parameters $h_3 = 2.06(0.27)$ and $\varsigma = 0.64(0.13)$; the consequences of these measurements are discussed in detail in Section~\ref{sec:postPK}; clearly the companion is a massive WD.

The relatively high orbital inclination determined in section~\ref{sec:postPK}, 
$i \, \sim \, 73.3^{\circ}$, implies that the pulse profile should be relatively
narrow. This is indeed observed (Figure~\ref{fig:prof3}).
This narrow profile yields precise timing. We measure a RMS residual of 5.18 $\mu$s; for this
reason the pulsar has been added to the NANOGrav PTA, which took a significant part of the
data used in this work. 

\subsection{PSR J0732$+$2314}
PSR J0732$+$2314 is an MSP with a spin period of 4.09 ms and a DM of 44.67 $\rm{pc}\, cm^{-3}$. It is in a 30.23-day ($e=0.000010$) orbit around a companion with a minimum companion mass of 0.15 $\rm{M_{\odot}}$ and a median companion mass of 0.18 $\rm{M_{\odot}}$.
Again, as in the case of PSR~J0509$+$0456, the minimum companion mass is low compared to the
\cite{TS99} prediction of $\sim \, 0.3 \, \rm{M_{\odot}}$ for the mass of a He WD that one should expect for this orbital period;
as for PSR~J0509$+$0456 this suggests that either the system
is being observed at a low orbital inclination, or, if it is being observed at a higher orbital inclination,
the pulsar mass is large. Similarly to PSR~J0509$+$0856, the lack of a detection of
the Shapiro delay and the characteristics of the pulse profile (Figure~\ref{fig:prof2}),
suggest a low inclination angle. A low inclination angle is consistent with our results
from \cite{Radhakrishnan+69} `rotating vector model' (RVM) on the L-band data for this pulsar.

The RVM fit yields the following parameters, the magnetic inclination angle $\alpha \, = \, 36.6^{+1.1}_{-1.3}\, \deg$
and the impact angle $\beta\, = \, 16.7_{-3.2}^{+2.2} \, \deg$. For the angle between the line of sight and the spin
axis ($\zeta = \alpha + \beta$) we get $\zeta \, = \, 53.3_{-2.5}^{+2.4}\, \deg$. Since this is likely aligned with
the orbital angular momentum \citep{Manchester+10}, then it indicates $i \, \sim \, 53.3^{+2.4}_{-2.5}\, \deg$.

Our timing solution has an RMS residual of 4.69 $\mu s$ and spans about 1.5 years. This pulsar follows the \citep{Phinney92} relationship between eccentricity and orbital period ($P_{b}$) evolution, see Figure~\ref{fig:Eccen_vs_Pb}. It fills a noticeable gap within the range of $P_{b} \sim 22 - 32$ days where the population is dominated
by eccentric binaries which have He WD companions \citep{2015ApJ...806..140K,2015ApJ...810...85C,Barr2017,2018A&A...612A..78O,2019ApJ...870...74S}. If we assume the
same for PSR~J0732$+$2314, this would mean that not all such systems in this
interval of orbital periods become eccentric.

This pulsar has also been included in the NANOGrav PTA. It could potentially be a very precise probe of the
Solar wind given its very low Ecliptic latitude ($\lambda_\odot\, =\, 111.188\, \deg$, $\beta_\odot\, = \, 1.496\, \deg$), with the Sun being in the vicinity of the pulsar in July 14. We have no data from near this epoch, so we
have not measured large DM variations caused by the Sun.

\subsection{PSR J0824$+$0028}
PSR J0824$+$0028 is an MSP with a spin period of 9.86 ms and a DM of 34.55 $\rm{pc}\, cm^{-3}$. It is in a 23.20-day, ($e=0.000230$) orbit around a companion with a minimum mass of 0.34 $\rm{M_{\odot}}$ and a median mass of 0.41 $\rm{M_{\odot}}$. Our timing observations span about 4.5 years and have a RMS residuals of 4.66 $\mu s$. A hint of Shapiro delay has been detected, with orthometric parameters $h_3 = 0.50(0.25)$ and $\varsigma = 0.85(0.11)$. There is no detection of the rate of advance of periastron, $\dot{\omega}$. Even though there is no $\dot{\omega}$, more observations at superior conjunction could lead to mass measurements from Shapiro delay alone. The mass of the companion is too large for a He WD; it is therefore more likely a CO WD. The orbital eccentricity of the system and the spin period of the pulsar match well the characteristics of other systems with CO WD companions; these have a wider range of eccentricities (generally they are more eccentric) than predicted by \cite{Phinney92}.

Similar to PSR~J0732$+$2314, its orbital period is also inside the range dominated
by the eccentric binaries mentioned above. However the companion for PSR J0824$+$0028
is not likely a He WD, so it certainly had a different evolution than the eccentric MSPs.

\subsection{PSR J2204$+$2700}
PSR J2204$+$2700 was the first binary pulsar found in AO327 data. It is an MRP with a DM of 35.07 $\rm{pc}\, cm^{-3}$. It is in a 815.24-day orbit, the seventh largest orbital period of any pulsar binary, and the widest found so far in AO327 data. Using the mass function, we get a minimum companion mass of 0.36 $\rm{M_{\odot}}$ and a median mass of 0.43 $\rm{M_{\odot}}$.
This system is part of a small group of binary pulsars
(which includes PSRs~J0214$+$5222, J2016$+$1948, J0407$+$1607, J1711$-$4322 and possibly J1840$-$0643,
see \citealt{atnf})
with orbital periods between 512 and 937 days,
spin periods of tens of ms (84.70 ms in this case) and orbital eccentricities of the order of $10^{-3}$
($e=0.00152$ in this case), which nevertheless still follow the relation predicted
by \citet{Phinney92}. Such systems have been recycled by mass accretion from the companion,
but given the large orbital separations, this has happened only during the brief time when
the companion was in its giant phase, which explains the mild recycling.

\subsection{Nature of the Binary Companions}
None of the five binaries presented here have a detectable optical counterpart for their companions
in the online DSS2 optical survey red/blue filters, nor in the 2MASS survey. Also, none of the binaries
show any evidence of eclipses in their timing residuals. This suggests that none of the companions is
a main sequence (or at least an extended) star, implying by default that the companions are WD stars. 

\subsection{Search for pulsars in $\gamma$ rays}
Many MSPs are known to emit detectable $\gamma$-ray pulses. This fact motivated us to search for these pulsars in $\gamma$ rays using the \texttt{TEMPO2}~\emph{fermi} plugin\footnote{https://fermi.gsfc.nasa.gov/ssc/data/analysis/user/Fermi\_plug\_doc.pdf} to fold data from the \emph{Fermi} Large Area Telescope (LAT) \citep{Fermi_mission} at the position of each pulsar. None of the 6 recycled pulsars were detected in $\gamma$-rays. The $\dot{E}$ values for four pulsars (J0154$+$1833, J0509$+$0856, J0732$+$2314 and J0824$+$0028) are above $2 \, \times \, 10^{33} \, \rm erg \, s^{-1}$. Several MSPs have been found with similar values of $\dot{E}$~\citep{Ransom+11,Abdo+09}. 

\begin{figure*}
\begin{center}
\includegraphics[width=\textwidth, angle=0]{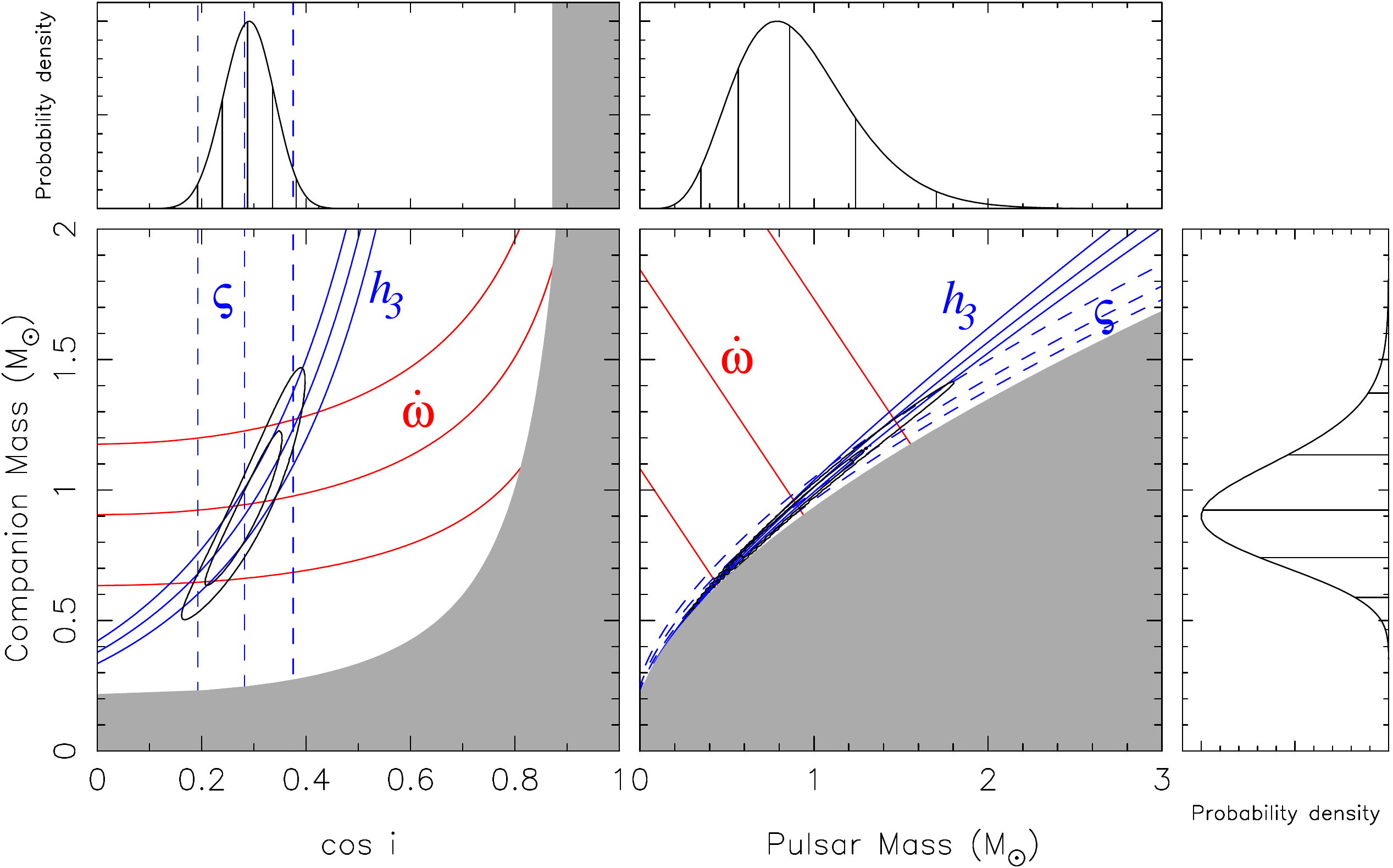}
\end{center}
\caption{Mass constraints for PSR~J10709+0458. In the lower plots, the
lines indicate the regions that are (according to general relativity)
consistent with the nominal and $\pm\, 1 \sigma$ measurements of $h_3$ (solid blue),
$\varsigma$ (dashed blue) and $\dot{\omega}$ (solid red) obtained from
the DDFWHE model (see Table~\ref{timingSol}).
The contour plots include 68.23 and 95.44\% of the total
2-dimensional probability density function (PDF), derived from the quality
($\chi^2$) of the {\sc tempo} fits using a DDGR model
to the ToA data set we have obtained for this pulsar.
The location of the regions of high probability is well described by the
$h_3$ and $\varsigma$ parameters and their uncertainties, with an important influence from $\dot{\omega}$: note that the distribution of
$\cos i$ is narrower than suggested by the measurement of $\varsigma$. 
In the left plot, we display
the cosine of the orbital inclination ($\cos i$, which has, for randomly
inclined orbits, a flat PDF) versus the companion mass ($M_{\rm{c}}$); the gray
region is excluded because the pulsar mass ($M_{\rm{p}}$) must be larger than 0.
In the right plot, we display $M_{\rm{p}}$ versus $M_{\rm{c}}$; the gray region is excluded
by the constraint $\sin i\, \leq\, 1$. The side panels display the 1-d PDFs
for $\cos i$ (top left), $M_{\rm{p}}$ (top right) and $M_{\rm{c}}$ (right). The vertical lines
in these PDFs indicate the median and the percentiles corresponding to
1 and 2 $\sigma$ around the median.}
\label{figure:mass_mass}
\end{figure*}

\section{Post-Keplerian parameters}\label{sec:postPK}

\subsection{Mass measurement for PSR~J0709+0458}

For PSR~J0709+0458, we have measured three model-independent Post-Keplerian (PK) parameters using the
DDFWHE model: $\dot{\omega} \, = \, 0.026(8) \, ^{\circ} \rm \, yr^{-1}$ and the ``orthometric'' Shapiro
delay parameters $h_3 \, = \, 1.86(21) \, \mu \rm s$ and $\varsigma \, = \, 0.75(7)$.
These measurements are only possible given the high timing precision 
of the system and the large companion mass: indeed, an orbital eccentricity of 0.00022 would generally
make the measurement of $\dot{\omega}$ impossible. 

The mass and inclination constraints introduced by these
parameters (assuming GR) are depicted graphically in 
Figure~\ref{figure:mass_mass}. Since we have three PK parameters
in this system, this results in a test of that theory. As we can see, all
PK parameters agree on the same masses, so GR passes the test posed by the measurement of
these 3 parameters. However, this implies no new constraints on alternative
theories of gravity given the limited  precision of the PK parameters.

To estimate the mass of the pulsar ($M_{\rm p}$), the companion ($M_{\rm c}$) and the
orbital inclination, $i$, we used the
Bayesian method described in detail by \cite{Barr2017} and
references therein. In this method, we make a map of the $\chi^2$ of
the solutions in a plane that has $\cos i$ (which has a constant
probability for randomly oriented orbits) and
$h_3$\footnote{We chose the latter variable instead of $M_{\rm{c}}$, used by
\cite{Barr2017}, in order to optimally cover
all the regions of the plane with relatively good fits to the timing data (i.e., with
a low $\chi^2$), thus spending  less time on regions of the parameter space with extremely
poor fits, which have negligible associated probability.} as axes.
For each point in this $\cos i$ - $h_3$ plane, we calculate the corresponding
$M_{\rm c}$ and total binary mass ($M$) and keep them fixed in a DDGR solution,
which accounts for all GR effects for these masses in a self-consistent way. 
This solution is then fitted to the TOAs, and the quality of the fit (as described by
the value of the $\chi^2$) is recorded for each point.

We then transform these $\chi^2$ maps into
a 2-D probability density function (PDF) in the $\cos i$ - $M_{\rm{c}}$
and $M_{\rm{p}}$ - $M_{\rm{c}}$ planes. The contours holding 68.23\% and 95.44\% of all
probability in these planes are displayed in 
Figure~\ref{figure:mass_mass}.
Projecting these 2-D PDFs onto the different axes, we obtain the 
probabilities for the masses and orbital inclination.

The regions of high probability in the main plots are well described by $h_3$,
$\varsigma$ and $\dot{\omega}$ parameters of the DDFWHE model in Table~\ref{timingSol}
and their uncertainties, as shown in figure~\ref{figure:mass_mass}.
For the main quantities, the 68.3 \% confidence limits are: 
$M_{\rm{c}} \, = \, 0.92_{-0.18}^{+0.21} \, \rm{M_{\odot}}$,
$M_{\rm{p}} \, = \, 0.86_{-0.29}^{+0.38} \, \rm{M_{\odot}}$ and
$i \, = 73(3)^{\circ}$; the 95.4\% confidence limits are:
$M_{\rm{c}} \, = \, 0.92_{-0.34}^{+0.45} \, \rm{M_{\odot}}$,
$M_{\rm{p}} \, = \, 0.86_{-0.51}^{+0.84} \, \rm{M_{\odot}}$ and
$i \, = 73(6)^{\circ}$.

The uncertainties are so large that these values are not yet astrophysically
useful (the pulsar mass can be almost anywhere
between 0 and 2 $\rm{M_{\odot}}$). However, this situation will
change with time: The measurement of $\dot{\omega}$ is already
constraining the masses significantly; indeed, the $\cos i$ distribution
is narrower than suggested by the measurement of $\varsigma$,
which implies that the masses have already a significantly narrower
distribution than we would obtain from the measurement of $\varsigma$
and $h_3$ alone.

As the timing baseline $T$ increases, the uncertainty of $\dot{\omega}$
will decrease proportionally to $T^{-3/2}$ (while for $\varsigma$ and
$h_3$, the uncertainties will
decrease with $T^{-1/2}$, which implies that the area of the $M_c$ - $\cos i$
plane allowed by the Shapiro delay measurement will decrease with $T^{-1}$).
As we can see in Figure~\ref{figure:mass_mass},
a more precise $\dot{\omega}$ will, in combination with the Shapiro delay,
yield much more precise masses. 

\section{Conclusions} \label{sec:conclusion}
Recycled pulsars are the major motivation for many large scale pulsar surveys; they present unique studies in physical applications and evolutionary studies of the end phase of binaries. Here we presented the follow-up and timing for six recycled pulsars discovered in the AO327 pulsar survey. Five of the recycled pulsars were found to be in a binary system, with three of them having He WD companions and two with more massive WD companions. Three of these pulsars (PSRs J0509+0856, J0709+0458 and J0732+2314) are being used in PTAs in efforts to detect low-frequency GWs. The AO327 pulsar survey keeps discovering pulsars that have good timing precision: the short exposure times (60 s) guarantees that any new
pulsars will be relatively bright. 

We were able to measure post-Keplerian parameters for two of our new systems, PSR~J0709+0458 and J0824+0028,
and for the first system we were able to determine the component masses, albeit at low precision.
Continued timing observations should substantially improve these PK parameters and, at least in the
first case, determine precise component masses in the not too distant future.

\begin{center}
    Acknowledgements 
\end{center}
This work would not have been possible without the high sensitivity of the Arecibo 305-m radio telescope, the professionalism and dedication of its excellent staff, and the capabilities of the PUPPI back-end developed by NRAO. J.G.M. was supported for this research through a stipend from the International Max Planck Research School (IMPRS) for Astronomy and Astrophysics at the University of Bonn and Cologne and acknowledges partial support through the Bonn Cologne Graduate School (BCGS) Honors Branch of Physics and Astronomy. J.S.D. was supported by the NASA Fermi program. P.G., K.S., M.A.M. are supported by the NANOGrav Physics Frontiers Center (award number \# 1430284). P.G. and M.A.M. are also supported by NSF award number OIA-1458952. G.D. is supported by the ERC synergy grant "BlackHoleCam: Imaging the Event Horizon of Black Holes" (Grant No. 610058).

\software{FITORBIT, PSRCHIVE (Hotan et al. 2004; van Straten et al. 2012), PRESTO (Ransom 2011), TEMPO (Nice et al. 2015), TEMPO2 (Hobbs, Edwards \& Manchester 2006), DRACULA (Freire \& Ridolfi 2018)}


\begin{deluxetable*}{lccc}
\tablecaption{Timing parameters for isolated pulsar and binary pulsars with ELL1H orbital model\label{timingSol_1}
}
\tablewidth{5000pt}
\tabletypesize{\footnotesize}
\tablehead{ \colhead{PSR } & \colhead{J0154+1833} & \colhead{J0509+0856}  & \colhead{J0732+2314} } 
\startdata
  \multicolumn{4}{c}{Observation and data reduction parameters}\\
  \hline
  Timing epoch (MJD) \dotfill & 56900 & 57384 & 58000 \\
  Span of timing data (MJD) \dotfill & 56550-57809 & 56516-58251 & 57522-58265 \\
  Number of TOAs \dotfill & 504 & 307 & 478\\
  Integration time per TOA (min) \dotfill & 5 & 5 & 5 \\
  TOA error scale factor (search/fold) \dotfill & $-$ & 2.04/1.5 & $-$/1.76\\
  rms post-fit residuals ($\mu \rm{s}$) \dotfill & 1.18 & 3.44 & 4.41 \\
  \hline
     \multicolumn{4}{c}{Spin and astrometric parameters}\\
  \hline
Right Ascension, $\alpha$ (J2000) \dotfill & 01:54:36.88273(3) & 05:09:22.23485(1) & 07:32:37.5156(8)  \\
Declination, $\delta$ (J2000) \dotfill     & +18:33:50.758(9)   & +08:56:25.0248(8)  & +23:14:54.21(7) \\
Proper Motion Right Ascension, $\mu_{\alpha} (\rm{mas \,  yr^{-1}})$  \dotfill & 10.3(9) & 5.4(2) & $-$  \\
Proper Motion Declination, $\mu_{\delta} (\rm{mas \, yr^{-1}})$\dotfill     & $-$8.9(1.9) & $-$4.3(5) & $-$ \\ 
Spin frequency, $\nu\, (\rm{Hz})$ \dotfill        & 422.90991368608(8) & 246.55815713199(3) & 244.4905830450 \\
Spin frequency derivativee, $\dot{\nu}$ ($10^{-15}$~Hz s$^{-1}$) \dotfill & $-$0.522(2) & $-$0.2682(7) & $-$0.36(2) \\
Dispersion Measure, DM (pc $\rm{cm^{-3}}$ ) \dotfill & 19.7978(1) & 38.318(4) & 44.6724(1) \\
Dispersion Measure derivative, DM1 (pc $\rm{cm^{-3}}$ $\rm{yr^{-1}}$) \dotfill & 0.0013(1) & 0.0006(2) & $-$0.0052(9) \\
Dispersion Measure derivative, DM2 (pc $\rm{cm^{-3}}$ $\rm{yr^{-2}}$) \dotfill &$-$&$-$&0.058(7) \\
Rotation measure, (rad m$^{−2}$) \dotfill & 21.6(1) & 42.4(6) & 3.6(1.4)\\
\hline
  \multicolumn{4}{c}{Binary Parameters}\\
  \hline
   Orbital Period, $P_b$ (days) \dotfill & $-$ & 4.907976893(1)  & 30.2300773(5) \\
   Projected Semi-major Axis, $x$ (lt-s) \dotfill & $-$ & 2.4580255(3) &  10.625842(2)\\
   Time of ascending node, $T_{\rm asc}$ (MJD) \dotfill & $-$ & 56519.1715605(3) & 57747.116391(7) \\
   $\epsilon_1 \, \equiv \, e \sin \omega$ \dotfill & $-$ &  0.0000118(4) & 0.0000086(2)\\
   $\epsilon_2 \, \equiv \, e \cos \omega$ \dotfill & $-$ & 0.0000190(3)  & 0.0000034(4)\\
   Orthometric amplitude of the Shapiro delay $h_3$ ($\mu$s) \dotfill & $-$ & 0.5(4) & $-$1.0(8)\\
   Orthometric amplitude of the Shapiro delay $h_4$ ($\mu$s) \dotfill & $-$ & 0.2(4) & $-$1.7(6) \\
  \hline
  \multicolumn{4}{c}{Derived parameters}\\
  \hline
  Galactic Longitude, $l$ (deg) \dotfill & 143.18 &  192.48 & 195.93 \\
  Galactic Latitude, $b$ (deg) \dotfill & $-$41.80 & $-$17.93 & 19.03\\
  DM Derived Distance, $d_{1}(NE2001)/d_{2}(YMW)$ (kpc) \dotfill & 0.86/1.62 & 1.45/0.82  & 1.66/1.15 \\
  Galactic height, $z_{1}/z_{2}$ (kpc) \dotfill &  $-$0.57/$-$1.08 & $-$0.45/$-$0.25 & 0.54/0.37 \\
  Transversal velocity, $v_{\perp}$(km s$^{-1}$) \dotfill & 56(14) & 48(2) & $-$  \\
  Spin period, $P\,  (\rm{s})$ \dotfill        & 0.0023645697763005(4) & 0.0040558382315642(6) & 0.004090137082357(4) \\
  Spin period derivative, $\dot{P}$ ($10^{-21}$~ss$^{-1}$) \dotfill & 2.92(1) & 4.41(1) & 6.1(3) \\
  Correction to $\dot{P}$ from Galactic acceleration, $P a /c$ ($10^{-21}$ss$^{-1} $) \dotfill & $-$0.21$^{+0.01}_{-0.2}$ & 0.20(5) & 0.19(6)   \\
  Correction to $\dot{P}$ from Shklovskii effect, $P \mu^2 / c$ ($10^{-21}$ss$^{-1} $) \dotfill & 0.7(2)  & 0.7(2) & $-$  \\
  Intrinsic $\dot{P}$, $\dot{P}_{\rm int} (10^{-21} \rm{s s^{-1}})$\dotfill & 2.2$^{+0.2}_{-0.2}$ & 3.5$^{+0.2}_{-0.2}$ & 5.9$^{+0.06}_{-0.06}$  \\
  Surface Magnetic Field Strength $B_0$ ($10^{9}$ G) \dotfill & 0.08 & 0.13 & 0.15 \\
  Characteristic Age, $\tau_c$ (Gyr) \dotfill & 12.8 & 14.5 & 10.6 \\
  Spin-down Luminosity, $\dot{E}$ (erg $\rm{s^{-1}}$) \dotfill & $8.7 \times 10^{33}$ & $2.6 \times 10^{33}$ & $3.5 \times 10^{33}$ \\
  Orbital Eccentricity, $e$ \dotfill & $-$ & 0000224(3) & 0.0000093(2)  \\
  Longitude of Periastron, $\omega$ ($^\circ$) \dotfill & $-$ & 31.9(9) & 68(3) \\
  Epoch of Periastron, $T_0$ (MJD) \dotfill & $-$ & 56519.60(1) & 57752.8(2) \\
  Mass Function, $f$ ($\rm{M_{\odot}}$) \dotfill & $-$ & 0006619679(3) & 0.0014095937(4)  \\
  Min. Companion Mass, $M_c$ ($\rm{M_{\odot}}$) \dotfill & $-$ & 0.11  & 0.15 \\
  \hline
  \multicolumn{4}{l}{In this and the next table, the TOAs measurements used the UTC(NIST) timescale and the parameters are given in Barycentric Dynamic Time}\\
  \multicolumn{4}{l}{(TDB). We used the DE 421 Solar System ephemeris.}\\
 \enddata
\end{deluxetable*}

\begin{deluxetable*}{lccc}
\tablecaption{Timing Parameters for binary pulsars with DDFWHE orbital model \label{timingSol}}
\tablewidth{5000pt}
\tabletypesize{\footnotesize}
\tablehead{\colhead{PSR } &  \colhead{J0709+0458} & \colhead{J0824+0028} & \colhead{J2204+2700} } 
\startdata
  \multicolumn{4}{c}{Observation and data reduction parameters}\\
  \hline
  Timing epoch (MJD) \dotfill & 56983 & 56600 & 56805\\
  Span of timing data (MJD) \dotfill & 56600-58328 & 56499-58213 & 55338-58327 \\
  Number of TOAs \dotfill & 1518 & 824 & 295\\
  Integration time per TOA (min) \dotfill & 5 & 5 & 5 \\
  TOA error scale factor (search/fold) \dotfill & 1.2/1.1 & $-$/$1.15$ & 1.04/1.07 \\
  rms post-fit residuals ($\mu \rm{s}$) \dotfill & 5.01 & 4.66 & 79.94\\
  \hline
    \multicolumn{4}{c}{Spin and astrometric parameters}\\
  \hline
Right Ascension, $\alpha$ (J2000) \dotfill & 07:09:08.36539(2) & 08:24:24.84022(5) & 22:04:43.609(3) \\
Declination, $\delta$ (J2000) \dotfill     & +04:58:51.4941(6) & 00:28:0.567(2) & +27:00:54.69(4)\\
Proper Motion Right Ascension, $\mu_{\alpha} (\rm{mas \,  yr^{-1}})$  \dotfill & $-$0.7(2) & $-$4.3(4) & 2(16) \\
Proper Motion Declination, $\mu_{\delta} (\rm{mas\,  yr^{-1}})$\dotfill     & $-$1.3(4)  & $-$9.2(1.3) & $-$3(3) \\ 
Spin frequency, $\nu\, (\rm{Hz})$ \dotfill        & 29.045300004132(1) & 101.40600778116(1) & 11.80601087935(2) \\
Spin frequency derivative, $\dot{\nu}$ ($10^{-15}$~Hz s$^{-1}$) \dotfill & $-$0.32092(3) & $-$1.5105(2) & $-$0.0182(3) \\
Dispersion Measure, DM (pc $\rm{cm^{-3}}$ ) \dotfill & 44.2664(5) & 34.5485(6) & 35.074(1) \\
Dispersion Measure derivative, DM1 (pc $\rm{cm^{-3}}$ $\rm{yr^{-1}}$) \dotfill & $-$0.0004(1)  & $-$  & $-$\\
Dispersion Measure derivative, DM2 (pc $\rm{cm^{-3}}$ $\rm{yr^{-2}}$) \dotfill &0.0008(1) & $-$&$-$ \\
Rotation measure, (rad m$^{−2}$) \dotfill &  43.4(3) & 39.3(5) & $-$ \\
\hline
  \multicolumn{4}{c}{Binary Parameters}\\
  \hline
   Orbital Period, $P_b$ (days) \dotfill & 4.366681(1) & 23.206955708(5) & 815.24544(5) \\
   Projected Semi-major Axis, $x$ (lt-s) \dotfill & 15.716583(5) & 18.988927(2) &210.68062(4) \\
   Epoch of Periastron, $T_0$ (MJD) \dotfill & 56983.4410(3) & 56519.296(1) &56635.37(3) \\
   Orbital Eccentricity, $e$ \dotfill & 0.00022539(7) & 0.00023072(5) & 0.0015227(8) \\
   Longitude of Periastron, $\omega$ ($^\circ$) \dotfill & 322.70(3) & 46.32(2) & 6.40(1) \\
   Rate of advance of periastron, $\dot{\omega}$ ($^\circ \, \rm yr^{-1}$) & 0.03(1) & $-$ & $-$ \\
   Orthometric amplitude of the Shapiro delay $h_3$ ($\mu$s) \dotfill & 1.8(2) & 0.5(2) & $-$ \\
   Orthometric ratio of the Shapiro delay $\varsigma$ \dotfill & 0.74(8) & 0.8(1) & $-$ \\
  \hline
  \multicolumn{4}{c}{Derived parameters}\\
  \hline
  Galactic Longitude, $l$ (deg) \dotfill & 210.49 & 223.57 & 82.99\\
  Galactic Latitude, $b$ (deg) \dotfill & 6.20 & 20.78 & $-$22.64\\
  DM Derived Distance, $d_{1}(NE2001)/d_{2}(YMW)$ (kpc) \dotfill & 1.79/1.20 & 1.53/1.68 & 2.15/3.15 \\
  Galactic height, $z_{1}/z_{2}$ (kpc) \dotfill & 0.19/0.13  & 0.54/0.59 & $-$0.82/$-$1.22\\
  Transversal velocity, $v_{\perp}$(km s$^{-1}$) \dotfill & 12(3) & 74(18) & 50(12) \\
  Spin period, $P \, (\rm{s})$ \dotfill        &  0.0344289781774587(9) & 0.009861348670368(1) & 0.0847026154913(2)\\
  Spin period derivative, $\dot{P}$ ($10^{-21}$~ss$^{-1}$) \dotfill & 380.35(1)  & 146.88(1) &  130.9(20) \\
  Correction to $\dot{P}$ from Galactic acceleration, $P a / c$ ($10^{-21}$ss$^{-1} $) \dotfill & 2.25$^{+0.51}_{-0.55}$ & $-$0.369$^{+0.008}_{-0.01}$ & $-$19(4) \\
  Correction to $\dot{P}$ from Shklovskii effect $P \mu^2 / c$ ($10^{-21}$ss$^{-1} $) \dotfill &  0.32(7) & 3.8(9) & 8(2) \\
  Intrinsic $\dot{P}$, $\dot{P}_{\rm int} (10^{-21} \rm{s s^{-1}})$\dotfill & 377.7$^{+0.6}_{-0.6}$ & 143.4$^{+0.9}_{-0.9}$ & 141.2$^{+1.7}_{-1.9}$ \\
  Surface Magnetic Field Strength $B_0$ ($10^{9}$ G) \dotfill & 3.64 & 1.21 & 3.35\\
  Characteristic Age, $\tau_c$ (Gyr) \dotfill & 1.4 & 1.0 & 10.3\\
  Spin-down Luminosity, $\dot{E}$ (erg $\rm{s^{-1}}$) \dotfill & $3.7 \times 10^{32}$ & $6.0 \times 10^{33}$ & $8.4 \times 10^{30} $ \\
  Mass Function, $f$ ($\rm{M_{\odot}}$ ) \dotfill & 0.21860219(1) & 0.0136504619(9)  & 0.0151070557(9) \\
  Min. Companion Mass, $M_c$ ($\rm{M_{\odot}}$) \dotfill &  1.11  & 0.34 & 0.36\\
  \hline
\enddata
\end{deluxetable*}



\begin{thebibliography}{}
\expandafter\ifx\csname natexlab\endcsname\relax\def\natexlab#1{#1}\fi
\providecommand{\url}[1]{\href{#1}{#1}}

\bibitem[{{Abdo} {et~al.}(2009){Abdo}, {Ackermann}, {Atwood}, {Axelsson},
  {Baldini}, {Ballet}, {Barbiellini}, {Bastieri}, {Battelino}, {Baughman},
  {Bechtol}, {Bellazzini}, {Berenji}, {Bloom}, {Bonamente}, {Borgland},
  {Bregeon}, {Brez}, {Brigida}, {Bruel}, {Burnett}, {Caliandro}, {Cameron},
  {Caraveo}, {Casandjian}, {Cecchi}, {Charles}, {Chekhtman}, {Cheung},
  {Chiang}, {Ciprini}, {Claus}, {Cognard}, {Cohen-Tanugi}, {Cominsky},
  {Conrad}, {Cutini}, {Dermer}, {de Angelis}, {de Palma}, {Digel}, {Dormody},
  {Silva}, {Drell}, {Dubois}, {Dumora}, {Farnier}, {Favuzzi}, {Focke},
  {Frailis}, {Fukazawa}, {Funk}, {Fusco}, {Gargano}, {Gasparrini}, {Gehrels},
  {Germani}, {Giebels}, {Giglietto}, {Giordano}, {Glanzman}, {Godfrey},
  {Grenier}, {Grondin}, {Grove}, {Guillemot}, {Guiriec}, {Hanabata}, {Harding},
  {Hayashida}, {Hays}, {Hughes}, {J{\'o}hannesson}, {Johnson}, {Johnson},
  {Johnson}, {Johnson}, {Kamae}, {Katagiri}, {Kataoka}, {Kawai}, {Kerr},
  {Kn{\"o}dlseder}, {Kocian}, {Komin}, {Kuehn}, {Kuss}, {Lande}, {Latronico},
  {Lee}, {Lemoine-Goumard}, {Longo}, {Loparco}, {Lott}, {Lovellette},
  {Lubrano}, {Madejski}, {Makeev}, {Marelli}, {Mazziotta}, {McConville},
  {McEnery}, {Meurer}, {Michelson}, {Mitthumsiri}, {Mizuno}, {Moiseev},
  {Monte}, {Monzani}, {Morselli}, {Moskalenko}, {Murgia}, {Nolan}, {Nuss},
  {Ohsugi}, {Omodei}, {Orlando}, {Ormes}, {Pancrazi}, {Paneque}, {Panetta},
  {Parent}, {Pepe}, {Pesce-Rollins}, {Piron}, {Porter}, {Rain{\`o}}, {Rando},
  {Razzano}, {Reimer}, {Reimer}, {Reposeur}, {Ritz}, {Rochester}, {Rodriguez},
  {Romani}, {Ryde}, {Sadrozinski}, {Sanchez}, {Sander}, {Parkinson},
  {Sgr{\`o}}, {Siskind}, {Smith}, {Smith}, {Spandre}, {Spinelli}, {Starck},
  {Strickman}, {Suson}, {Tajima}, {Takahashi}, {Tanaka}, {Thayer}, {Thayer},
  {Theureau}, {Thompson}, {Tibaldo}, {Torres}, {Tosti}, {Tramacere},
  {Uchiyama}, {Usher}, {Van Etten}, {Vilchez}, {Vitale}, {Waite}, {Watters},
  {Webb}, {Wood}, {Ylinen}, \& {Ziegler}}]{Abdo+09}
{Abdo}, A.~A., {Ackermann}, M., {Atwood}, W.~B., {et~al.} 2009, \apj, 699, 1171

\bibitem[{{Alpar} {et~al.}(1982){Alpar}, {Cheng}, {Ruderman}, \&
  {Shaham}}]{Alpar1982}
{Alpar}, M.~A., {Cheng}, A.~F., {Ruderman}, M.~A., \& {Shaham}, J. 1982, \nat,
  300, 728

\bibitem[{{Antoniadis}(2014)}]{Antoniadis14_formationEccenMSP}
{Antoniadis}, J. 2014, \apjl, 797, L24

\bibitem[{{Antoniadis} {et~al.}(2016){Antoniadis}, {Kaplan}, {Stovall},
  {Freire}, {Deneva}, {Koester}, {Jenet}, \& {Martinez}}]{J2234_optical}
{Antoniadis}, J., {Kaplan}, D.~L., {Stovall}, K., {et~al.} 2016, \apj, 830, 36

\bibitem[{{Antoniadis} {et~al.}(2013){Antoniadis}, {Freire}, {Wex}, {Tauris},
  {Lynch}, {van Kerkwijk}, {Kramer}, {Bassa}, {Dhillon}, {Driebe}, {Hessels},
  {Kaspi}, {Kondratiev}, {Langer}, {Marsh}, {McLaughlin}, {Pennucci}, {Ransom},
  {Stairs}, {van Leeuwen}, {Verbiest}, \& {Whelan}}]{Antoniadis+13_J0348}
{Antoniadis}, J., {Freire}, P.~C.~C., {Wex}, N., {et~al.} 2013, Science, 340,
  448

\bibitem[{{Archibald} {et~al.}(2018){Archibald}, {Gusinskaia}, {Hessels},
  {Deller}, {Kaplan}, {Lorimer}, {Lynch}, {Ransom}, \&
  {Stairs}}]{Archibald+18_TripleSystem}
{Archibald}, A.~M., {Gusinskaia}, N.~V., {Hessels}, J.~W.~T., {et~al.} 2018,
  \nat, 559, 73

\bibitem[{{Arzoumanian} {et~al.}(2018{\natexlab{a}}){Arzoumanian}, {Baker},
  {Brazier}, {Burke-Spolaor}, {Chamberlin}, {Chatterjee}, {Christy}, {Cordes},
  {Cornish}, {Crawford}, {Thankful Cromartie}, {Crowter}, {DeCesar},
  {Demorest}, {Dolch}, {Ellis}, {Ferdman}, {Ferrara}, {Folkner}, {Fonseca},
  {Garver-Daniels}, {Gentile}, {Haas}, {Hazboun}, {Huerta}, {Islo}, {Jones},
  {Jones}, {Kaplan}, {Kaspi}, {Lam}, {Lazio}, {Levin}, {Lommen}, {Lorimer},
  {Luo}, {Lynch}, {Madison}, {McLaughlin}, {McWilliams}, {Mingarelli}, {Ng},
  {Nice}, {Park}, {Pennucci}, {Pol}, {Ransom}, {Ray}, {Rasskazov}, {Siemens},
  {Simon}, {Spiewak}, {Stairs}, {Stinebring}, {Stovall}, {Swiggum}, {Taylor},
  {Vallisneri}, {van Haasteren}, {Vigeland}, {Zhu}, \& {The NANOGrav
  Collaboration}}]{GWB_limit_NANOGrav}
{Arzoumanian}, Z., {Baker}, P.~T., {Brazier}, A., {et~al.} 2018{\natexlab{a}},
  \apj, 859, 47

\bibitem[{{Arzoumanian} {et~al.}(2018{\natexlab{b}}){Arzoumanian}, {Baker},
  {Brazier}, {Burke-Spolaor}, {Chamberlin}, {Chatterjee}, {Christy}, {Cordes},
  {Cornish}, {Crawford}, {Thankful Cromartie}, {Crowter}, {DeCesar},
  {Demorest}, {Dolch}, {Ellis}, {Ferdman}, {Ferrara}, {Folkner}, {Fonseca},
  {Garver-Daniels}, {Gentile}, {Haas}, {Hazboun}, {Huerta}, {Islo}, {Jones},
  {Jones}, {Kaplan}, {Kaspi}, {Lam}, {Lazio}, {Levin}, {Lommen}, {Lorimer},
  {Luo}, {Lynch}, {Madison}, {McLaughlin}, {McWilliams}, {Mingarelli}, {Ng},
  {Nice}, {Park}, {Pennucci}, {Pol}, {Ransom}, {Ray}, {Rasskazov}, {Siemens},
  {Simon}, {Spiewak}, {Stairs}, {Stinebring}, {Stovall}, {Swiggum}, {Taylor},
  {Vallisneri}, {van Haasteren}, {Vigeland}, {Zhu}, \& {NANOGrav
  Collaboration}}]{NANOGRAV_11yr+18}
---. 2018{\natexlab{b}}, \apj, 859, 47

\bibitem[{{Atwood} {et~al.}(2009){Atwood}, {Abdo}, {Ackermann}, {Althouse},
  {Anderson}, {Axelsson}, {Baldini}, {Ballet}, {Band}, {Barbiellini}, \&
  et~al.}]{Fermi_mission}
{Atwood}, W.~B., {Abdo}, A.~A., {Ackermann}, M., {et~al.} 2009, \apj, 697, 1071

\bibitem[{{Barr} {et~al.}(2017){Barr}, {Freire}, {Kramer}, {Champion},
  {Berezina}, {Bassa}, {Lyne}, \& {Stappers}}]{Barr2017}
{Barr}, E.~D., {Freire}, P.~C.~C., {Kramer}, M., {et~al.} 2017, \mnras, 465,
  1711

\bibitem[{{Bhattacharya} \& {van den
  Heuvel}(1991)}]{Bhattacharya&van_den_Heuvel1991}
{Bhattacharya}, D., \& {van den Heuvel}, E.~P.~J. 1991, \physrep, 203, 1

\bibitem[{{Cameron} {et~al.}(2018){Cameron}, {Champion}, {Kramer}, {Bailes},
  {Barr}, {Bassa}, {Bhandari}, {Bhat}, {Burgay}, {Burke-Spolaor}, {Eatough},
  {Flynn}, {Freire}, {Jameson}, {Johnston}, {Karuppusamy}, {Keith}, {Levin},
  {Lorimer}, {Lyne}, {McLaughlin}, {Ng}, {Petroff}, {Possenti}, {Ridolfi},
  {Stappers}, {van Straten}, {Tauris}, {Tiburzi}, \& {Wex}}]{Cameron_J1757}
{Cameron}, A.~D., {Champion}, D.~J., {Kramer}, M., {et~al.} 2018, \mnras, 475,
  L57

\bibitem[{{Camilo} {et~al.}(2015){Camilo}, {Kerr}, {Ray}, {Ransom},
  {Sarkissian}, {Cromartie}, {Johnston}, {Reynolds}, {Wolff}, {Freire},
  {Bhattacharyya}, {Ferrara}, {Keith}, {Michelson}, {Saz Parkinson}, \&
  {Wood}}]{2015ApJ...810...85C}
{Camilo}, F., {Kerr}, M., {Ray}, P.~S., {et~al.} 2015, \apj, 810, 85

\bibitem[{{Cordes} \& {Lazio}(2002)}]{CordesLazio2002}
{Cordes}, J.~M., \& {Lazio}, T.~J.~W. 2002, ArXiv Astrophysics e-prints,
  astro-ph/0207156

\bibitem[{{Damour} \& {Deruelle}(1986)}]{Damour86}
{Damour}, T., \& {Deruelle}, N. 1986, Ann.~Inst.~Henri Poincar{\'e}
  Phys.~Th{\'e}or., Vol.~44, No.~3, p.~263 - 292, 44, 263

\bibitem[{{Damour} \& {Taylor}(1991)}]{1991ApJ...366..501D}
{Damour}, T., \& {Taylor}, J.~H. 1991, \apj, 366, 501

\bibitem[{{Deneva} {et~al.}(2013){Deneva}, {Stovall}, {McLaughlin}, {Bates},
  {Freire}, {Martinez}, {Jenet}, \& {Bagchi}}]{Deneva13}
{Deneva}, J.~S., {Stovall}, K., {McLaughlin}, M.~A., {et~al.} 2013, \apj, 775,
  51

\bibitem[{{Freire} \& {Ridolfi}(2018)}]{connect}
{Freire}, P. C.~C., \& {Ridolfi}, A. 2018, \mnras, 476, 4794

\bibitem[{{Freire} \& {Wex}(2010)}]{FreireWex}
{Freire}, P.~C.~C., \& {Wex}, N. 2010, \mnras, 409, 199

\bibitem[{{Freire} {et~al.}(2012){Freire}, {Wex}, {Esposito-Far{\`e}se},
  {Verbiest}, {Bailes}, {Jacoby}, {Kramer}, {Stairs}, {Antoniadis}, \&
  {Janssen}}]{Freire+12}
{Freire}, P.~C.~C., {Wex}, N., {Esposito-Far{\`e}se}, G., {et~al.} 2012,
  \mnras, 423, 3328

\bibitem[{{Gravity Collaboration} {et~al.}(2018){Gravity Collaboration},
  {Abuter}, {Amorim}, {Anugu}, {Baub{\"o}ck}, {Benisty}, {Berger}, {Blind},
  {Bonnet}, {Brandner}, {Buron}, {Collin}, {Chapron}, {Cl{\'e}net}, {Coud{\'e}
  Du Foresto}, {de Zeeuw}, {Deen}, {Delplancke-Str{\"o}bele}, {Dembet},
  {Dexter}, {Duvert}, {Eckart}, {Eisenhauer}, {Finger}, {F{\"o}rster
  Schreiber}, {F{\'e}dou}, {Garcia}, {Garcia Lopez}, {Gao}, {Gendron},
  {Genzel}, {Gillessen}, {Gordo}, {Habibi}, {Haubois}, {Haug}, {Hau{\ss}mann},
  {Henning}, {Hippler}, {Horrobin}, {Hubert}, {Hubin}, {Jimenez Rosales},
  {Jochum}, {Jocou}, {Kaufer}, {Kellner}, {Kendrew}, {Kervella}, {Kok},
  {Kulas}, {Lacour}, {Lapeyr{\`e}re}, {Lazareff}, {Le Bouquin}, {L{\'e}na},
  {Lippa}, {Lenzen}, {M{\'e}rand}, {M{\"u}ler}, {Neumann}, {Ott}, {Palanca},
  {Paumard}, {Pasquini}, {Perraut}, {Perrin}, {Pfuhl}, {Plewa}, {Rabien},
  {Ram{\'{\i}}rez}, {Ramos}, {Rau}, {Rodr{\'{\i}}guez-Coira}, {Rohloff},
  {Rousset}, {Sanchez-Bermudez}, {Scheithauer}, {Sch{\"o}ller}, {Schuler},
  {Spyromilio}, {Straub}, {Straubmeier}, {Sturm}, {Tacconi}, {Tristram},
  {Vincent}, {von Fellenberg}, {Wank}, {Waisberg}, {Widmann}, {Wieprecht},
  {Wiest}, {Wiezorrek}, {Woillez}, {Yazici}, {Ziegler}, \&
  {Zins}}]{2018A&A...615L..15G}
{Gravity Collaboration}, {Abuter}, R., {Amorim}, A., {et~al.} 2018, \aap, 615,
  L15

\bibitem[{{Hotan} {et~al.}(2004){Hotan}, {van Straten}, \&
  {Manchester}}]{PSRCHIVE}
{Hotan}, A.~W., {van Straten}, W., \& {Manchester}, R.~N. 2004, \pasa, 21, 302

\bibitem[{{Knispel} {et~al.}(2015){Knispel}, {Lyne}, {Stappers}, {Freire},
  {Lazarus}, {Allen}, {Aulbert}, {Bock}, {Bogdanov}, {Brazier}, {Camilo},
  {Cardoso}, {Chatterjee}, {Cordes}, {Crawford}, {Deneva}, {Eggenstein},
  {Fehrmann}, {Ferdman}, {Hessels}, {Jenet}, {Karako-Argaman}, {Kaspi}, {van
  Leeuwen}, {Lorimer}, {Lynch}, {Machenschalk}, {Madsen}, {McLaughlin},
  {Patel}, {Ransom}, {Scholz}, {Siemens}, {Spitler}, {Stairs}, {Stovall},
  {Swiggum}, {Venkataraman}, {Wharton}, \& {Zhu}}]{2015ApJ...806..140K}
{Knispel}, B., {Lyne}, A.~G., {Stappers}, B.~W., {et~al.} 2015, \apj, 806, 140

\bibitem[{{Kramer} {et~al.}(2006){Kramer}, {Stairs}, {Manchester},
  {McLaughlin}, {Lyne}, {Ferdman}, {Burgay}, {Lorimer}, {Possenti}, {D'Amico},
  {Sarkissian}, {Hobbs}, {Reynolds}, {Freire}, \& {Camilo}}]{GRTestsdoublePSR}
{Kramer}, M., {Stairs}, I.~H., {Manchester}, R.~N., {et~al.} 2006, Science,
  314, 97

\bibitem[{{Lange} {et~al.}(2001){Lange}, {Camilo}, {Wex}, {Kramer}, {Backer},
  {Lyne}, \& {Doroshenko}}]{ELL1}
{Lange}, C., {Camilo}, F., {Wex}, N., {et~al.} 2001, \mnras, 326, 274

\bibitem[{{Lazaridis} {et~al.}(2009){Lazaridis}, {Wex}, {Jessner}, {Kramer},
  {Stappers}, {Janssen}, {Desvignes}, {Purver}, {Cognard}, {Theureau}, {Lyne},
  {Jordan}, \& {Zensus}}]{2009MNRAS.400..805L}
{Lazaridis}, K., {Wex}, N., {Jessner}, A., {et~al.} 2009, \mnras, 400, 805

\bibitem[{{Lentati} {et~al.}(2015){Lentati}, {Taylor}, {Mingarelli}, {Sesana},
  {Sanidas}, {Vecchio}, {Caballero}, {Lee}, {van Haasteren}, {Babak}, {Bassa},
  {Brem}, {Burgay}, {Champion}, {Cognard}, {Desvignes}, {Gair}, {Guillemot},
  {Hessels}, {Janssen}, {Karuppusamy}, {Kramer}, {Lassus}, {Lazarus}, {Liu},
  {Os{\l}owski}, {Perrodin}, {Petiteau}, {Possenti}, {Purver}, {Rosado},
  {Smits}, {Stappers}, {Theureau}, {Tiburzi}, \& {Verbiest}}]{GWB_limit_EPTA}
{Lentati}, L., {Taylor}, S.~R., {Mingarelli}, C.~M.~F., {et~al.} 2015, \mnras,
  453, 2576

\bibitem[{{Manchester} {et~al.}(2005){Manchester}, {Hobbs}, {Teoh}, \&
  {Hobbs}}]{atnf}
{Manchester}, R.~N., {Hobbs}, G.~B., {Teoh}, A., \& {Hobbs}, M. 2005, \aj, 129,
  1993

\bibitem[{{Manchester} {et~al.}(2010){Manchester}, {Kramer}, {Stairs},
  {Burgay}, {Camilo}, {Hobbs}, {Lorimer}, {Lyne}, {McLaughlin}, {McPhee},
  {Possenti}, {Reynolds}, \& {van Straten}}]{Manchester+10}
{Manchester}, R.~N., {Kramer}, M., {Stairs}, I.~H., {et~al.} 2010, \apj, 710,
  1694

\bibitem[{{Martinez} {et~al.}(2015){Martinez}, {Stovall}, {Freire}, {Deneva},
  {Jenet}, {McLaughlin}, {Bagchi}, {Bates}, \& {Ridolfi}}]{Martinez_J0453+1559}
{Martinez}, J.~G., {Stovall}, K., {Freire}, P.~C.~C., {et~al.} 2015, \apj, 812,
  143

\bibitem[{{Martinez} {et~al.}(2017){Martinez}, {Stovall}, {Freire}, {Deneva},
  {Tauris}, {Ridolfi}, {Wex}, {Jenet}, {McLaughlin}, \&
  {Bagchi}}]{Martinez_J1411+2551}
---. 2017, \apjl, 851, L29

\bibitem[{{McGaugh}(2018)}]{2018arXiv180809435M}
{McGaugh}, S. 2018, ArXiv e-prints, arXiv:1808.09435

\bibitem[{{Octau} {et~al.}(2018){Octau}, {Cognard}, {Guillemot}, {Tauris},
  {Freire}, {Desvignes}, \& {Theureau}}]{2018A&A...612A..78O}
{Octau}, F., {Cognard}, I., {Guillemot}, L., {et~al.} 2018, \aap, 612, A78

\bibitem[{{Phinney}(1992)}]{Phinney92}
{Phinney}, E.~S. 1992, Philosophical Transactions of the Royal Society of
  London Series A, 341, 39

\bibitem[{{Radhakrishnan} \& {Cooke}(1969)}]{Radhakrishnan+69}
{Radhakrishnan}, V., \& {Cooke}, D.~J. 1969, \aplett, 3, 225

\bibitem[{{Ransom} {et~al.}(2011){Ransom}, {Ray}, {Camilo}, {Roberts}, {{\c
  C}elik}, {Wolff}, {Cheung}, {Kerr}, {Pennucci}, {DeCesar}, {Cognard}, {Lyne},
  {Stappers}, {Freire}, {Grove}, {Abdo}, {Desvignes}, {Donato}, {Ferrara},
  {Gehrels}, {Guillemot}, {Gwon}, {Harding}, {Johnston}, {Keith}, {Kramer},
  {Michelson}, {Parent}, {Saz Parkinson}, {Romani}, {Smith}, {Theureau},
  {Thompson}, {Weltevrede}, {Wood}, \& {Ziegler}}]{Ransom+11}
{Ransom}, S.~M., {Ray}, P.~S., {Camilo}, F., {et~al.} 2011, \apjl, 727, L16

\bibitem[{{Ransom} {et~al.}(2014){Ransom}, {Stairs}, {Archibald}, {Hessels},
  {Kaplan}, {van Kerkwijk}, {Boyles}, {Deller}, {Chatterjee},
  {Schechtman-Rook}, {Berndsen}, {Lynch}, {Lorimer}, {Karako-Argaman}, {Kaspi},
  {Kondratiev}, {McLaughlin}, {van Leeuwen}, {Rosen}, {Roberts}, \&
  {Stovall}}]{Ransom+14_TripleSystem}
{Ransom}, S.~M., {Stairs}, I.~H., {Archibald}, A.~M., {et~al.} 2014, \nat, 505,
  520

\bibitem[{{Shannon} {et~al.}(2015){Shannon}, {Ravi}, {Lentati}, {Lasky},
  {Hobbs}, {Kerr}, {Manchester}, {Coles}, {Levin}, {Bailes}, {Bhat},
  {Burke-Spolaor}, {Dai}, {Keith}, {Os{\l}owski}, {Reardon}, {van Straten},
  {Toomey}, {Wang}, {Wen}, {Wyithe}, \& {Zhu}}]{GWB_limit_PPTA}
{Shannon}, R.~M., {Ravi}, V., {Lentati}, L.~T., {et~al.} 2015, Science, 349,
  1522

\bibitem[{{Shklovskii}(1970)}]{Shklovskii70}
{Shklovskii}, I.~S. 1970, \sovast, 13, 562

\bibitem[{{Stovall} {et~al.}(2018){Stovall}, {Freire}, {Chatterjee},
  {Demorest}, {Lorimer}, {McLaughlin}, {Pol}, {van Leeuwen}, {Wharton},
  {Allen}, {Boyce}, {Brazier}, {Caballero}, {Camilo}, {Camuccio}, {Cordes},
  {Crawford}, {Deneva}, {Ferdman}, {Hessels}, {Jenet}, {Kaspi}, {Knispel},
  {Lazarus}, {Lynch}, {Parent}, {Patel}, {Pleunis}, {Ransom}, {Scholz},
  {Seymour}, {Siemens}, {Stairs}, {Swiggum}, \& {Zhu}}]{Stovall_J1946}
{Stovall}, K., {Freire}, P.~C.~C., {Chatterjee}, S., {et~al.} 2018, \apjl, 854,
  L22

\bibitem[{{Stovall} {et~al.}(2019){Stovall}, {Freire}, {Antoniadis}, {Bagchi},
  {Deneva}, {Garver-Daniels}, {Martinez}, {McLaughlin}, {Arzoumanian},
  {Blumer}, {Brook}, {Cromartie}, {Demorest}, {DeCesar}, {Dolch}, {Ellis},
  {Ferdman}, {Ferrara}, {Fonseca}, {Gentile}, {Jones}, {Lam}, {Lorimer},
  {Lynch}, {Ng}, {Nice}, {Pennucci}, {Ransom}, {Spiewak}, {Stairs}, {Swiggum},
  {Vigeland}, \& {Zhu}}]{2019ApJ...870...74S}
{Stovall}, K., {Freire}, P.~C.~C., {Antoniadis}, J., {et~al.} 2019, \apj, 870,
  74

\bibitem[{{Tauris} {et~al.}(2011){Tauris}, {Langer}, \&
  {Kramer}}]{TaurisLangerKramer2011}
{Tauris}, T.~M., {Langer}, N., \& {Kramer}, M. 2011, \mnras, 416, 2130

\bibitem[{{Tauris} {et~al.}(2012){Tauris}, {Langer}, \&
  {Kramer}}]{TaurisLangerKramer2012}
---. 2012, \mnras, 425, 1601

\bibitem[{{Tauris} {et~al.}(2015){Tauris}, {Langer}, \&
  {Podsiadlowski}}]{Tauris_Ultra-stripSN}
{Tauris}, T.~M., {Langer}, N., \& {Podsiadlowski}, P. 2015, \mnras, 451, 2123

\bibitem[{{Tauris} \& {Savonije}(1999)}]{TS99}
{Tauris}, T.~M., \& {Savonije}, G.~J. 1999, \aap, 350, 928

\bibitem[{{Tauris} \& {van den Heuvel}(2006)}]{Tauris_van_den_Heuvel2006}
{Tauris}, T.~M., \& {van den Heuvel}, E.~P.~J. 2006, {Formation and evolution
  of compact stellar X-ray sources}, ed. W.~H.~G. {Lewin} \& M.~{van der Klis},
  623--665

\bibitem[{{Taylor}(1992)}]{TaylorRG}
{Taylor}, J.~H. 1992, Royal Society of London Philosophical Transactions Series
  A, 341, 117

\bibitem[{{van Straten} {et~al.}(2012){van Straten}, {Demorest}, \&
  {Oslowski}}]{PSRCHIVE2012}
{van Straten}, W., {Demorest}, P., \& {Oslowski}, S. 2012, Astronomical
  Research and Technology, 9, 237

\bibitem[{{Weisberg} \& {Huang}(2016)}]{2016ApJ...829...55W}
{Weisberg}, J.~M., \& {Huang}, Y. 2016, \apj, 829, 55

\bibitem[{{Yao} {et~al.}(2017){Yao}, {Manchester}, \& {Wang}}]{YMW_model2017}
{Yao}, J.~M., {Manchester}, R.~N., \& {Wang}, N. 2017, \apj, 835, 29

\end{thebibliography}
\end{document}